\documentclass[notitlepage]{revtex4-1}

\usepackage[utf8]{inputenc}
\usepackage[colorlinks=true,citecolor=blue,linkcolor=blue]{hyperref}
\usepackage[normalem]{ulem}
\usepackage{url}
\usepackage{graphicx,wrapfig,float,slashed,cancel}
\usepackage{amsmath,amssymb,epsfig,graphicx,xcolor,stmaryrd}
\usepackage{bm}
\usepackage{enumitem}

\definecolor{darkblue}{RGB}{1, 90, 173}

\begin{document}


\title{Determination of the possible quantum numbers for the newly observed $\Xi_b(6227)^0$ state}

\author{K.~Azizi}
\email{ kazem.azizi@ut.ac.ir}
\thanks{Corresponding author}
\affiliation{Department of Physics, University of Tehran, North Karegar Avenue, Tehran
14395-547, Iran}
\affiliation{Department of Physics, Do\u gu\c s University,
Ac{\i}badem-Kad{\i}k\"oy, 34722 Istanbul, Turkey}
\affiliation{School of Particles and Accelerators, Institute for Research in Fundamental Sciences (IPM) P.O. Box 19395-5531, Tehran, Iran}
\author{Y.~Sarac}
\email{yasemin.sarac@atilim.edu.tr}
\affiliation{Electrical and Electronics Engineering Department,
Atilim University, 06836 Ankara, Turkey}
\author{H.~Sundu}
\email{ hayriye.sundu@kocaeli.edu.tr}
\affiliation{Department of Physics, Kocaeli University, 41380 Izmit, Turkey}

\date{\today}

\preprint{}

\begin{abstract}
	The LHCb Collaboration recently reported the observation of a new excited bottom baryon $\Xi_b(6227)^0$ and announced an improvement in the measurements related to the  previously observed $\Xi_b(6227)^-$ state. We conduct an analysis for $\Xi_b(6227)^0$ state considering it as  isospin partner of the $\Xi_b(6227)^-$ resonance and possibly $1P$ or $2S$ excited state with spin $J=\frac{3}{2}$. The corresponding masses for  both possibilities have consistent results with  the experimental data, indicating that only with the mass sum rules, one can not make exact decision on the nature and quantum numbers of this state. To go further, the decays of these possible excited states to $\Xi_b^- \pi^+$ final state are also considered and the relevant strong coupling constants are extracted from the light cone sum rules. The obtained decay width values support the possibility of $\Xi_b(6227)^0$ to be the $1P$ excited state of $\Xi_b(5945)^0$ baryon.   
\end{abstract}


\maketitle

\renewcommand{\thefootnote}{\#\arabic{footnote}}
\setcounter{footnote}{0}
\section{\label{sec:level1}Introduction}\label{intro} 

The progress achieved in the experimental investigations has resulted in many observations recently. Among these observations are the excited states of the heavy baryons with a single $b$ or $c$ quark. These observations have produced a subsequent revival of interest in the theoretical investigations of these excited states. The contribution of the investigations of heavy baryons for the understanding of the strong interaction has an important place. With the presence of one heavy quark, they provide an attractive ground to study the dynamics of light quarks. These studies provide a way not only to improve our understanding of the confinement mechanism but also a test for the quark model and heavy quark symmetry. Their investigations gain importance also from the point of view that by the result of these investigations, one may check the predictions obtained by different theoretical works on their interactions and internal organizations and provide insights into the experimental researches. 

The quark model predicted the existence of various excited states of heavy baryons such as their orbital or radial excitations. While  almost all the ground states of the heavy baryon were observed, the number of the observed excited states of bottom baryons is limited~\cite{Aaij:2012da,Chatrchyan:2012ni,Aaij:2014yka,Aaij:2015yoy,Aaij:2018yqz}. In the last few years some observations~\cite{Aaij:2020cex,Aaij:2018tnn,Aaij:2019amv} were reported indicating the observed states to be in favor of excited states of the bottom baryons. The excited states of heavy baryons were investigated  via various theoretical models and some of these investigations dates back to even before their observations. Among this theoretical models were the various quark models~\cite{Capstick:1986bm,Ebert:2007nw,Garcilazo:2007eh,Roberts:2007ni,Valcarce:2008dr,Ebert:2011kk,Yoshida:2015tia,Karliner:2015ema,Thakkar:2016dna,Shah:2016mig,Migura:2006ep} searching their mass spectrum and decay mechanisms ~\cite{Hussain:1999sp,Ivanov:1998wj,Ivanov:1999bk,Albertus:2005zy,Zhong:2007gp,Hernandez:2011tx,Liu:2012sj,Chen:2016iyi,Wang:2017kfr,Nagahiro:2016nsx,Yao:2018jmc,Wang:2018fjm,Chen:2018vuc,Wang:2019uaj}. Their masses and decays were investigated via QCD sum rule method~\cite{Zhu:2000py,Wang:2010it,Mao:2015gya,Chen:2016phw,Wang:2017vtv,Mao:2017wbz,Aliev:2018lcs,Cui:2019dzj,Azizi:2020tgh}, light cone QCD sum rule method~\cite{Zhu:1998ih,Wang:2009cd,Wang:2009ic,Aliev:2009jt,Aliev:2010yx,Aliev:2012tb,Aliev:2014bma,Aliev:2016xvq,Chen:2017sci,Agaev:2017ywp,Aliev:2018vye}, ${}^3P_{0}$ model~\cite{Chen:2007xf,Ye:2017yvl,Ye:2017dra,Chen:2017aqm,Yang:2018lzg,Guo:2019ytq,Liang:2019aag,Lu:2019rtg}, lattice QCD~\cite{Padmanath:2013bla,Bahtiyar:2015sga,Bali:2015lka,Bahtiyar:2016dom} and bound state picture~\cite{Chow:1995nw}. For more details about excited heavy baryon states see also~\cite{Klempt:2009pi,Crede:2013kia,Chen:2016spr} and the references therein. 

In the last decade, we witnessed the observation of various excites bottom baryon states. The $\Lambda_b^{*}(5912){}^0$, $\Lambda_b^{*}(5920){}^0$~\cite{Aaij:2012da}, $\Xi_b'(5935)^-$, $\Xi_b(5955)^-$~\cite{Aaij:2014yka}, $\Xi_b(6227)^-$~\cite{Aaij:2018yqz}, $\Sigma_b(6097)^{\pm}$~\cite{Aaij:2018tnn}, $\Lambda_b(6146)^0$, $\Lambda_b(6152)^0$~\cite{Aaij:2019amv}, $\Omega_b(6316)^-$, $\Omega_b(6330)^-$, $\Omega_b(6340)^-$ and $\Omega_b(6350)^-$~\cite{Aaij:2020cex} are among these states. The LHCb's measurements of the mass and width for $\Xi_b(6227)^-$~\cite{Aaij:2018yqz} were improved recently in acompany with a new observation~\cite{Aaij:2020fxj}. The new resonance represented as $\Xi_b(6227)^0$ was reported to have the following resonance parameters~\cite{Aaij:2020fxj}:
\begin{eqnarray}
m_{\Xi_b(6227)^0}=6227.1^{+1.4}_{-1.5}\pm 0.5~\mathrm{MeV},~~~~~~~~
\Gamma(\Xi_b(6227)^0)=18.6^{+5.0}_{-4.1}\pm 1.4~\mathrm{MeV}.\nonumber\\
\end{eqnarray}          
On the other hand the improved measurements of the $\Xi_b(6227)^-$ were reported as~\cite{Aaij:2020fxj}:
\begin{eqnarray}
m_{\Xi_b(6227)^-}=6227.9\pm 0.8\pm 0.5~\mathrm{MeV},~~~~~~~~
\Gamma(\Xi_b(6227)^-)=19.9\pm 2.1\pm 1.5~\mathrm{MeV}.\nonumber\\
\end{eqnarray} 

After the first observation of $\Xi_b(6227)^-$~\cite{Aaij:2018yqz}, to enlighten the quantum numbers of the state, it was investigated via various theoretical models. In Ref.~\cite{Wang:2018fjm} its strong decay was considered in constituent quark model, and possible spin parity assignment was suggested as $J^P=\frac{3}{2}^-$ or $J^P=\frac{5}{2}^-$. Its two-body strong decay investigation indicted its being a candidate of the $P$-wave state with $J^P=\frac{3}{2}^-$ or $J^P=\frac{5}{2}^-$~\cite{Chen:2018orb}. QCD sum rules calculation given in Ref.~\cite{Aliev:2018lcs} supported its being $1P$ angular-orbital excited state of the $\Xi_b(5955)^-$ baryon with quantum numbers $J^P=\frac{3}{2}^-$. The Ref.~\cite{Cui:2019dzj} analyzed the state via the QCD sum rule method within the framework of heavy quark effective theory and interpreted its quantum numbers as $J^P=\frac{3}{2}^-$. Refs.~\cite{Yang:2019cvw,Yang:2020zrh} applied the light cone QCD sum rules to study its decays to ground-state bottom baryons and a meson considering the heavy quark effective theory. The consistency of the quantum numbers of the $\Xi_b(6227)^-$ with a $P$-wave state with $J^P=\frac{1}{2}^-$, $\frac{3}{2}^-$ and $\frac{5}{2}^-$  was also indicated by mass calculation in the quark-diquark picture~\cite{Faustov:2020gun,Faustov:2018vgl}. In another work made in the quark-diquark picture, the preferable spin of this particle was stated to be $J^P=\frac{3}{2}^-$~\cite{Jia:2019bkr}. This state was also interpreted considering the molecular point of view~\cite{Yu:2018yxl,Huang:2018bed,Nieves:2019jhp,Zhu:2020lza}. These analyses indicate that, to distinguish different interpretations, we need more investigations providing a deeper understanding of the substructure and quantum numbers of the considered resonance.

After the first report of the $\Xi_b(6227)^-$~\cite{Aaij:2018yqz}, we performed a QCD sum rule analyses for this state considering different possible scenarios for its quantum numbers such as $1P/2S$ excitations of the $\Xi_b^-$ and $\Xi_b'(5935)^-$ baryons with $J^P=\frac{1}{2}$ or $1P/2S$ excitations of the $\Xi_b(5955)^-$ with $J^P=\frac{3}{2}$~\cite{Aliev:2018lcs}. For all considered cases, we calculated the corresponding masses using QCD sum rules. All obtained results showed consistency with the experimental one, which directed us to a further investigation via their possible decays to $\Lambda_b^0K^-$ and $\Xi_b^-\pi$. The obtained decay widths supported the $\Xi_b(6227)^-$ to be a $1P$ angular-orbital excited state of $\Xi_b(5955)^-$ with $J^P=\frac{3}{2}^-$. In this work, we consider the new observation of $\Xi_b(6227)^0$ and its interpretation as  an isospin partner of the $\Xi_b(6227)^-$ by the LHCb Collaboration~\cite{Aaij:2020fxj}.  By considering also our previous assignment of the $\Xi_b(6227)^-$~\cite{Aliev:2018lcs}, we take into account two possible quantum numbers for $\Xi_b(6227)^0$ state as $1P$ or $2S$  excited state of the ground state $\Xi_b(5945)^0$  with  $J=\frac{3}{2}$. We investigate their masses and decays into the ground states $\Xi_b(5945)^-$ and $\pi^+$ to shed light on the quantum numbers of $\Xi_b(6227)^0$. To this end, we apply two-point QCD sum rule~\cite{Shifman:1978bx,Shifman:1978by,Ioffe81} and its extension light cone QCD sum rule (LCSR)~\cite{Braun:1988qv,Balitsky:1989ry,Chernyak:1990ag}  for masses and the coupling constants, respectively. In calculations, a proper interpolating current for the state is considered based on the quark content of the state and  the considered quantum numbers. We aim to assign probable quantum numbers to the state based on the comparison of the obtained masses and decay widths with experimental values.

This article is structured as follows. Section~\ref{II} is set aside for the details of the QCD sum rule calculations for the mentioned spectroscopic parameters. In section~\ref{III} the sum rules for the transition of the considered candidate states to the $\Xi_b^-$ and $\pi^+$ final states are presented. The last section is devoted to a summary, which also gives a comparison of the obtained results with experimental data.

\section{QCD sum rule calculations for spectral properties}\label{II}

The spectral properties such as masses and current coupling constants for the considered candidates $1P$ and $2S$ excited states of $\Xi_b(5945)^0$ are calculated in this section. The calculation is performed via the following two-point correlation function:
\begin{equation}
\Pi _{\mu \nu}(k)=i\int d^{4}xe^{ik\cdot x}\langle 0|\mathcal{T}\{\eta_{\mu}(x){\bar{\eta}_{\nu}}(0)\}|0\rangle ,  \label{eq:CorrF1}
\end{equation}
where $k$ is the four-momentum of the particle and  $\eta_{\mu}$ represents the interpolating current of the considered states, and it carries the quantum numbers of the considered states. This interpolating current is written in terms of the quark fields  considering the quark content of the state. For the state of interest, the interpolating current has the following form:
\begin{eqnarray}
\label{Eq:Current1}
\eta_{\mu} = \sqrt{\frac{2}{3}} \epsilon^{abc} \Big\{ (u^{aT} C \gamma_\mu s^b) b^c + (s^{aT} C
\gamma_\mu b^b) u^c + (b^{aT} C \gamma_\mu u^b) s^c \Big\}~,
\end{eqnarray}
with  $a$, $b$, and $c$ corresponding to the color indices. The operator $C$ is the charge conjugation.  

The above correlator can either be calculated in terms of the hadronic degrees of freedom such as masses, current coupling constants or in terms of the QCD degrees of freedom such as quark-gluon condensates, the strong coupling constant, quark masses, etc. In the analyses, both ways of calculations are performed, and obtained results are matched considering the coefficients of the same Lorentz structures for both sides. These matches lead to the QCD sum rules for the interested physical quantities, which are masses and the corresponding current coupling constants for this section. 

To calculate the correlator in terms of hadronic degrees of freedom, we treat the interpolating currents as operators annihilating or creating the hadrons. To proceed in the calculations, we insert  complete sets of hadronic states between the interpolating currents that have the same quantum numbers as the current itself. This process is followed by integration over $x$, which results in  
 \begin{eqnarray}
\Pi_{\mu\nu}^{\mathrm{Had}}(k)&=&\frac{\langle 0|\eta_{\mu } |\Xi_b^0(k,s;\frac{3}{2})\rangle \langle \Xi_b^0(k,s;\frac{3}{2})|\bar{\eta}_{\nu}|0\rangle}{m^{2}-k^{2}}
+\frac{\langle 0|\eta_{\mu } |\widetilde{\Xi}_b^0(k,s;\frac{3}{2})\rangle \langle \widetilde{\Xi}_b^0(k,s;\frac{3}{2})|\bar{\eta}_{\nu}|0\rangle}{\widetilde{m}^{2}-k^{2}}
+\ldots,
\label{eq:phys1a}
\end{eqnarray}
for $1P$ excitation and
\begin{eqnarray}
\Pi_{\mu\nu}^{\mathrm{Had}}(k)&=&\frac{\langle 0|\eta_{\mu } |\Xi_b^0(k,s;\frac{3}{2})\rangle \langle \Xi_b^0(k,s;\frac{3}{2})|\bar{\eta}_{\nu}|0\rangle}{m^{2}-k^{2}}
+\frac{\langle 0|\eta_{\mu } |\Xi_b^0{}'(k,s;\frac{3}{2})\rangle \langle \Xi_b^0{}'(k,s;\frac{3}{2})|\bar{\eta}_{\nu}|0\rangle}{m'^2-k^{2}}
+\ldots,
\label{eq:phys1b}
\end{eqnarray}
for $2S$ excitation. Here we consider  the  possibilities for the state under study  as $1P$ and $2S$ separately and analyze the results accordingly. In the last results $m$, $\tilde{m}$ and $m'$ represent the masses of the spin-3/2 ground ($\Xi_b(5945)^{0}$ in PDG), $1P$ and $2S$ states, respectively. Here,  $|\Xi_b^0(k,s;\frac{3}{2})\rangle$, $|\widetilde{\Xi}_{b}^0(k,s;\frac{3}{2})\rangle$ and $|\Xi_{b}^{0}{}'(k,s;\frac{3}{2})\rangle$ also respectively denote the ground, $1P$ and $2S$ states with $J=\frac{3}{2}$  and dots are used for the contributions of the higher states and continuum. The matrix elements, $\langle 0|\eta_{\mu } |\Xi_b^0(k,s;\frac{3}{2})\rangle$, $\langle 0|\eta_{\mu } |\widetilde{\Xi}_b^0(k,s;\frac{3}{2})\rangle$ and $\langle 0|\eta_{\mu } |\Xi_b^0{}'(k,s;\frac{3}{2})\rangle$ are defined in terms of the current coupling constants, $\lambda$ , $\tilde{\lambda}$ and $\lambda'$ as
\begin{eqnarray}
\langle 0|\eta_{\mu } |\Xi_b^0(k,s;\frac{3}{2})\rangle &=&\lambda u_{\mu}(k,s),
\nonumber \\
\langle 0|\eta_{\mu } |\widetilde{\Xi}_{b}^0(k,s;\frac{3}{2})\rangle
 &=&\widetilde{\lambda}\gamma_5 u_{\mu}(k,s),
\nonumber \\
\langle 0|\eta_{\mu } |\Xi_{b}^{0}{}'(k,s;\frac{3}{2})\rangle
&=&\lambda'u_{\mu}(k,s),
\label{eq:Res2ek}
\end{eqnarray}
where $u_{\mu}(k,s)$ is the Rarita-Schwinger spinor. 

Here we should note  that the current $\eta_{\mu}$ couples not only to the obove mentioned spin--3/2 states, but also to the
spin--1/2 states of both parities. To get the contributions of the spin-3/2 states,  we should remove the unwanted pollution coming from the  spin-1/2 particles.
The general form of the matrix element of $\eta_{\mu}$ between the spin-1/2 and vacuum
states can be parameterize as
\begin{eqnarray}\label{EliminationSpin1half2}
{\langle}0|\eta_{\mu}|\frac{1}{2}(k,s){\rangle}=\left(C_1 k_{\mu}+C_2
\gamma_{\mu}\right)u(k,s),
\end{eqnarray}
where $C_1$ and $C_2$ are some constants.
Now, we multiply both sides of the above equation with $\gamma^\mu$, and use the
condition $\gamma^\mu \eta_{\mu}=0$, to get $C_1$ in terms of  $C_2$. Hence,
\begin{eqnarray}\label{EliminationSpin1half2-1}
{\langle}0|\eta_{\mu}|\frac{1}{2}^+(k,s){\rangle}=C_2\left(-\frac{4}{m_{\frac{1}{2}^+}}
k_{\mu}+ \gamma_{\mu}\right)u(k,s),
\end{eqnarray}
for the positive parity and
\begin{eqnarray}\label{EliminationSpin1half2-2}
{\langle}0|\eta_{\mu}|\frac{1}{2}^-(k,s){\rangle}=C_2\gamma_{5}\left(-\frac{4}{m_{\frac{1}{2}^-}}
k_{\mu}+ \gamma_{\mu}\right)u(k,s),
\end{eqnarray}
for the negative parity states are obtained.
From these equations we see  that the unwanted contributions corresponding to  the spin-1/2 states with both parities
 are proportional to either $k_\mu$ or $\gamma_\mu$.
 To remove these contributions, first we will order the  Dirac matrices as $\gamma_\mu \rlap/{k}
 \gamma_\nu$ and then set to zero  the terms with $\gamma_\mu$ in the beginning and $\gamma_\nu$ at the end and those terms which are proportional to $k_\mu$ or $k_\nu$. But before that, using these matrix elements and the summation over spins as
\begin{eqnarray}\label{Rarita}
\sum_s  u_{\mu} (k,s)  \bar{u}_{\nu} (k,s) &= &-(\!\not\!{k} + m)\Big[g_{\mu\nu} -\frac{1}{3} \gamma_{\mu} \gamma_{\nu} - \frac{2k_{\mu}k_{\nu}}{3m^{2}} +\frac{k_{\mu}\gamma_{\nu}-k_{\nu}\gamma_{\mu}}{3m} \Big],
\end{eqnarray}
in Eqs.~(\ref{eq:phys1a}) and (\ref{eq:phys1b}) the following results  are obtained:
\begin{eqnarray}\label{PhyssSide}
\Pi_{\mu\nu}^{\mathrm{Had}}(k)&=&-\frac{\lambda^{2}}{k^{2}-m^{2}}(\!\not\!{k} + m)\Big[g_{\mu\nu} -\frac{1}{3} \gamma_{\mu} \gamma_{\nu} - \frac{2k_{\mu}k_{\nu}}{3m^{2}} +\frac{k_{\mu}\gamma_{\nu}-k_{\nu}\gamma_{\mu}}{3m} \Big]\nonumber \\
&-&\frac{\widetilde{\lambda}^{2}}{k^{2}-\widetilde{m}^{2}}(\!\not\!{k} - \widetilde{m})\Big[g_{\mu\nu} -\frac{1}{3} \gamma_{\mu} \gamma_{\nu} - \frac{2k_{\mu}k_{\nu}}{3\widetilde{m}^{2}} +\frac{k_{\mu}\gamma_{\nu}-k_{\nu}\gamma_{\mu}}{3\widetilde{m}} \Big]+\ldots,
\end{eqnarray}
\begin{eqnarray}\label{PhyssSide}
\Pi_{\mu\nu}^{\mathrm{Had}}(k)&=&-\frac{\lambda^{2}}{k^{2}-m^{2}}(\!\not\!{k} + m)\Big[g_{\mu\nu} -\frac{1}{3} \gamma_{\mu} \gamma_{\nu} - \frac{2k_{\mu}k_{\nu}}{3m^2} +\frac{k_{\mu}\gamma_{\nu}-k_{\nu}\gamma_{\mu}}{3m} \Big]\nonumber\\&-&\frac{\lambda'{}^{2}}{k^{2}-m'{}^{2}}(\!\not\!{k} + m')\Big[g_{\mu\nu} -\frac{1}{3} \gamma_{\mu} \gamma_{\nu} - \frac{2k_{\mu}k_{\nu}}{3m'{}^{2}} +\frac{k_{\mu}\gamma_{\nu}-k_{\nu}\gamma_{\mu}}{3m'} \Big]+\ldots,
\end{eqnarray}
where the first and second equations represent the results for the $1S+ 1P $ and $ 1S+2S $ possiblities, respectively.

As for the calculation of the same correlator, Eq.~(\ref{eq:CorrF1}), in terms of QCD degrees of freedom, the interpolating current is placed inside it explicitly, and operator product expansion (OPE) is applied. First, the possible contractions between quark fields are performed, and the result is converted to the ones given in terms of the light and heavy quark propagators. The corresponding propagators in coordinate space are used, and via Fourier transformation, the calculations proceed in momentum space, and finally, the results in terms of QCD degrees of freedom are achieved.

The results obtained from the hadronic and QCD sides are matched considering the coefficients of the same Lorentz structures in both sides after the application of Borel transformation to both sides. Borel transformation ensures the suppression of the contributions coming from the continuum and higher states. The mentioned structures in the calculations are the $ \!\not\!{k}g_{\mu\nu}$ and $g_{\mu\nu}$. These structures are selected after the  ordering of the Dirac matrices as stated above and removing the unwanted spin-1/2 pollution.  The final results for QCD sum rules obtained from the matches of the coefficients of these structures are as follows:
\begin{eqnarray}
\lambda^{2} e^{-\frac{m^{2}}{M^{2}}}+\widetilde{\lambda}^2(\lambda'{}^{ 2}) e^{-\frac{\widetilde{m}^2(m'{}^{2})}{M^{2}}}&=&\Pi^{\mathrm{QCD}}_{1},
\nonumber \\
m \lambda^{2} e^{-\frac{m^{2}}{M^{2}}}\mp \widetilde{m}(m') \widetilde{\lambda}^{2}(\lambda'{}^{2}) e^{-\frac{\widetilde{m}^{2}(m'{}^2)}{M^{2}}}&=&\Pi^{\mathrm{QCD}}_{2},
\label{Eq:sumrule2}
\end{eqnarray}
where $-$ and $+$ signs in the last equation are used for the $ 1P $ excited $\widetilde{\Xi}_b^0$ and $ 2S $ excited $\Xi^0_b{}'$ states, respectively. The functions $\Pi^{\mathrm{QCD}}_{1,2}$  represent the Borel transformed coefficients of the structures $\!\not\!{k}g_{\mu\nu}$  and $g_{\mu\nu}$  in the QCD side. These functions are lengthy, hence, as an example, we only present the explicit expression of the $ \Pi^{\mathrm{QCD}}_{1} $ function in the Appendix.

\begin{figure}[h!]
\begin{center}
\includegraphics[totalheight=5cm,width=7cm]{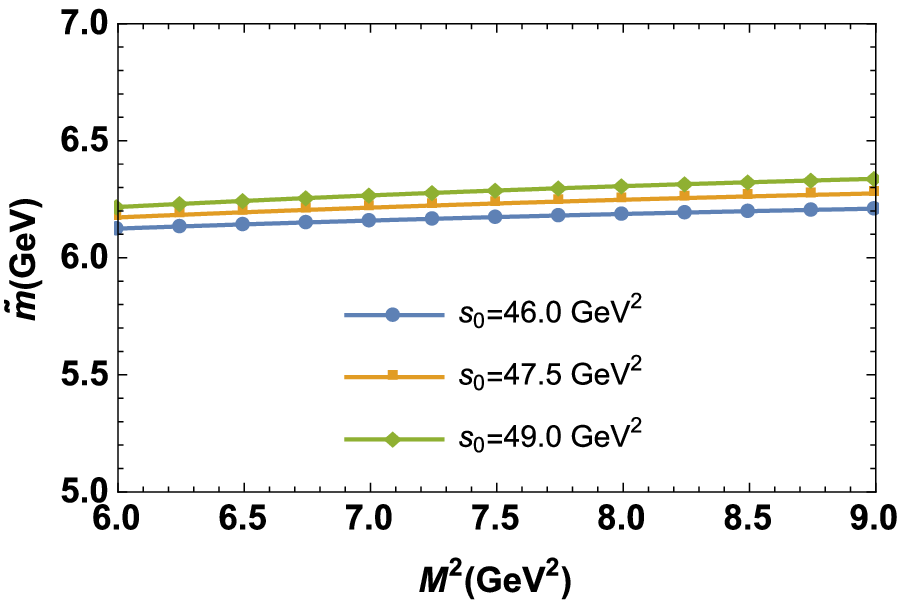}
\includegraphics[totalheight=5cm,width=7cm]{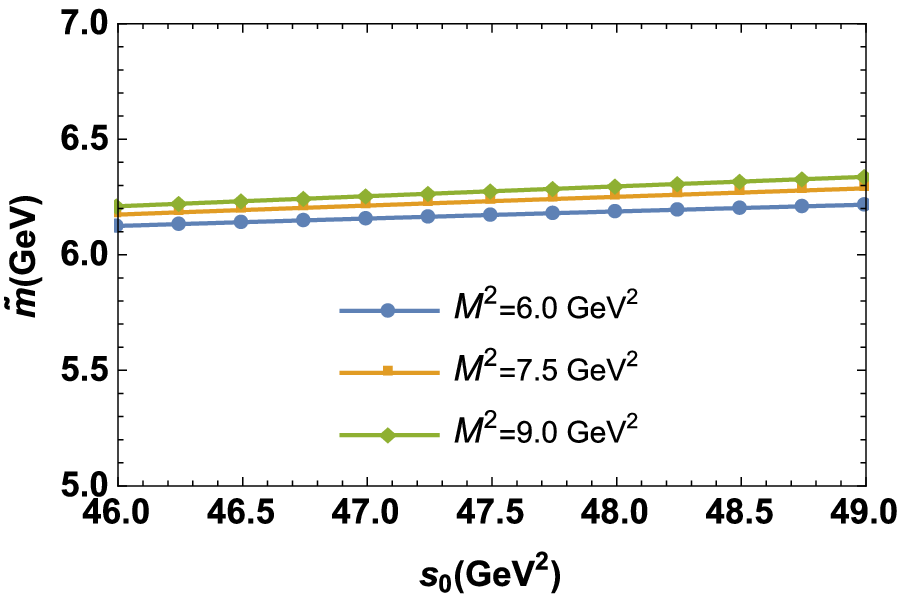}
\end{center}
\caption{\textbf{Left:} The mass $\widetilde{m}$ obtained for the  $ 1P $ excitation of $\Xi_b(5945)^{0}$  baryon vs Borel parameter $M^2$.
\textbf{Right:} The mass $\widetilde{m}$ obtained for the $ 1P $ excitation of $\Xi_b(5945)^{0}$  baryon  vs threshold parameter $s_0$.}
\label{gr:mXib}
\end{figure}

To  get the numerical results for quantities using the obtained sum rules, some input parameters are required, such as the mass of the ground state $\Xi_b(5945)^0$, quark masses, quark-gluon condensates, etc. The values of these quantities are gathered in Table~\ref{tab:Param}. Note that, in this table, we also present some other input parameters that are required in the next section. 
\begin{table}[tbp]
\begin{tabular}{|c|c|}
\hline\hline
Parameters & Values \\ \hline\hline
$ m_{\Xi_b(5945)^{0}}$                   & $5952.3\pm 0.6~\mathrm{MeV}$ \cite{Zyla:2020zbs}\\
$ m_{\Xi_b^{-}}$                   & $5797.0\pm 0.6~\mathrm{MeV}$ \cite{Zyla:2020zbs}\\
$m_{b}$                                  & $4.18^{+0.04}_{-0.03}~\mathrm{GeV}$ \cite{Zyla:2020zbs}\\
$m_{s}$                                  & $128^{+12}_{-4}~\mathrm{MeV}$ \cite{Zyla:2020zbs}\\
$m_{u}$                                  & $2.16^{+0.49}_{-0.26}~\mathrm{MeV}$ \cite{Zyla:2020zbs}\\
$ \lambda_{\Xi_b}$                       & $0.054\pm 0.012~\mathrm{GeV}^3$ \cite{Azizi:2016dmr}\\
$\langle \bar{q}q \rangle$ & $(-0.24\pm 0.01)^3$ $\mathrm{GeV}^3$ \cite{Belyaev:1982sa}  \\
$\langle \bar{s}s \rangle $              & $0.8\langle \bar{q}q \rangle$ \cite{Belyaev:1982sa} \\
$m_{0}^2 $                               & $(0.8\pm0.1)$ $\mathrm{GeV}^2$ \cite{Belyaev:1982sa}\\
$\langle g_s^2 G^2 \rangle $             & $4\pi^2 (0.012\pm0.004)$ $~\mathrm{GeV}
^4 $\cite{Belyaev:1982cd}\\
$ \Lambda $                              & $ (0.5\pm0.1) $ $\mathrm{GeV} $ \cite{Chetyrkin:2007vm} \\
\hline\hline
\end{tabular}%
\caption{Some input parameters used in the calculations of masses and coupling constants.}
\label{tab:Param}
\end{table}
In addition to the parameters given in Table~\ref{tab:Param}, there are two more auxiliary parameters required in the analyses. These are the threshold parameter $s_0$ and the Borel parameter $M^2$. The $s_0$ has a relation to the energy of the excited state with the same quantum numbers of state under question. However, due to lacking knowledge for these excited states, the corresponding working intervals are  determined considering the standard prescriptions of the method: namely the dominance of the first two resonances over other excited states and the convergence of OPE.  Besides, the weak dependencies of the results on the auxiliary  parameters  are applied. Following these criteria, the subsequent working intervals are obtained:
\begin{eqnarray}
46~\mbox{GeV}^2 \leq s_0 \leq 49~\mbox{GeV}^2,
\end{eqnarray}
and
\begin{eqnarray}
6~\mbox{GeV}^2 \leq M^2\leq 9~\mbox{GeV}^2.
\end{eqnarray}
Adopting these intervals for $s_0$ and $M^2$, the masses and the current coupling constants are obtained for the candidate $1P$ and $2S$ states as
\begin{eqnarray}
\tilde{m}=6225.47 \pm 106.29~\mathrm{MeV},~~~~~~\tilde{\lambda}=0.081\pm 0.010~\mathrm{GeV}^3,
\end{eqnarray}
and
\begin{eqnarray}
m'=6225.47 \pm 106.29~\mathrm{MeV},~~~~~~\lambda'=0.546\pm 0.039~\mathrm{GeV}^3,
\end{eqnarray}
respectively. Note that in these calculations, the ground state mass is used as given in Table~\ref{tab:Param} as an  input. The errors present in the results arise from the uncertainties of the input parameters and the auxiliary ones as well. To depict the dependencies of the results on the auxiliary parameters, we provide the mass plots of candidate $1P$ state as  functions of $M^2$ and $s_0$ in Fig.~\ref{gr:mXib}. This figure  shows the weak variations of the results  with respect to  the auxiliary parameters in their working intervals. 

These results indicate that the masses of both the $1P$ and $2S$ excitations are equal and consistent with the experimentally observed mass value of the $\Xi_b(6227)^0$ state. Therefore, to further deepen our prediction, we extend our analyses to the decays of these candidate states to $\Xi_b^-\pi^+$ final state. The details for this part of the investigation are presented in Section~\ref{III}.

\section{Transitions of the excited  $\Xi_b$ states to $\Xi_b^- \pi^+$}\label{III}

The mass sum rules performed for the experimentally observed $\Xi_b(6227)^0$ state in the previous section  led to  the same mass values for  candidates  $1P$ and $2S$ excitations of the $\Xi_b(5945)^0$ state,   but to different current coupling constants.  This prevented us to make an assignment on the nature of the discovered state. Therefore, to deepen the analyses and get the exact quantum number predictions, we provide further investigations that include the decays of the considered candidate states to the final states $\Xi_b^-$ and $ \pi^+$. Here $\Xi_b^-$ denotes the ground state spin-1/2 particle, which we will denote it by $ |\Xi_b^- (p,s;\frac{1}{2})\rangle $ in the following. In this section, the details of these analyses are presented. To compute the decay widths, we need to calculate the strong  coupling constants for the considered decays, and we shall calculate them using LCSR starting with the following correlation function
\begin{equation}
\Pi _{\mu}(p,q)=i\int d^{4}xe^{ip\cdot x}\langle \pi^+(q)|\mathcal{T}\{\eta_{\Xi_b^{-}
}(x)\bar{\eta}_{\mu}(0)\}|0\rangle,  \label{eq:CorrF2}
\end{equation}
where $ |\pi^+(q) \rangle$ in the correlator represents the on-shell $\pi$-meson state with momentum $q$, $p$ is the momentum of the final baryon and we assign the momentum $p'=p+q$ to the initial particle. In the calculations, the interpolating current $\eta_{\mu}$ from Eq.~(\ref{Eq:Current1}) for spin-$\frac{3}{2}$ initial state is used.    In this section,  both the possible cases $1P$ and $2S$  are considered for the initial state, as well. 
The spin-$\frac{1}{2}$ single heavy $\Xi_b^-$ baryon in the final state belongs to the $ \bar 3_F $  representation of  SU(3) flavor in the family of the single heavy bottom baryons (for more information see for instance Ref.~\cite{Aliev:2010yx}). 
The current interpolating the spin-$\frac{1}{2}$ $\Xi_b^-$ state has the following form:
\begin{eqnarray}
\eta_{\Xi_b^-} &=& {1\over \sqrt{6}} \epsilon^{abc} \Big\{ 2 (d^{aT} C
s^b) \gamma_5 b^c + (d^{aT} C b^b) \gamma_5 s^c +
(b^{aT} C s^b) \gamma_5 d^c + 2 \beta (d^{aT} C \gamma_5 s^b) b^c \nonumber\\&+& 
\beta (d^{aT} C \gamma_5 b^b) s^c +\beta
(b^{aT} C \gamma_5 s^b) d^c \Big\}~,\label{intfield12}
\end{eqnarray}
where $\beta$ is a mixing parameter. This parameter is used to combine the two possible current to make a general interpolating field. For more information about how the  current in Eq. (\ref{intfield12}) is obtained using all the quantum numbers of the $\Xi_Q$  ($Q=b$ or $c$), for instance,  see Ref. \cite{Aliev:2018ube}.  The working region for $\beta$, as an auxiliary parameter, is obtained demanding the condition that the physical quantities depend on it relatively weakly.  In order to scan all regions, we set $\beta=\tan \theta$ and plot the physical quantities in terms of  $\cos \theta$ in the interval $ [-1,1] $. Our analyses show that in the following working windows, the results weakly depend on the $\cos \theta$:
\begin{eqnarray}
-1\leq\cos\theta\leq -0.3 ~~~~~\mbox{and} ~~~~~~0.3\leq \cos\theta\leq 1~, 
\end{eqnarray}
which will be used in extraction of the numerical values for the quantities under study in this section.

To proceed in the calculations, we first insert complete sets of hadronic states having the same quantum numbers with the interpolating currents into the correlator to attain the hadronic sides as
\begin{eqnarray}
\Pi ^{\mathrm{Had}}_{\mu}(p,q)&=&\frac{\langle 0|\eta _{\Xi_b^-}|\Xi_b^- (p,s;\frac{1}{2})\rangle
}{p^{2}-m_{\Xi_b^-}^{2}}\langle \pi^+(q)\Xi_b^-(p,s;\frac{1}{2})|\Xi_b^{0}
(p^{\prime },s^{\prime };\frac{3}{2})\rangle  \frac{\langle \Xi_b^{0}(p^{\prime },s^{\prime
};\frac{3}{2})|\bar{\eta}_{\mu}|0\rangle }{p^{\prime
2}-m^{2}}\nonumber\\
&+&\frac{\langle 0|\eta _{\Xi_b^-}|\Xi_b^- (p,s;\frac{1}{2})\rangle
}{p^{2}-m_{\Xi_b^-}^{2}}\langle \pi^+(q)\Xi_b^-(p,s;\frac{1}{2})|\widetilde{\Xi}_b^0
(p^{\prime },s^{\prime };\frac{3}{2})\rangle  \frac{\langle \widetilde{\Xi}_b^{0}(p^{\prime },s^{\prime
};\frac{3}{2})|\bar{\eta}_{\mu}|0\rangle }{p^{\prime
2}-\widetilde{m}^{2}}
 +\ldots ,  \label{eq:SRDecay2a}
\end{eqnarray} 
when we consider the $1P$ state, and
\begin{eqnarray}
\Pi ^{\mathrm{Had}}_{\mu}(p,q)&=&\frac{\langle 0|\eta _{\Xi_b^-}|\Xi_b^- (p,s;\frac{1}{2})\rangle
}{p^{2}-m_{\Xi_b^-}^{2}}\langle \pi^+(q) \Xi_b^-(p,s;\frac{1}{2})|\Xi_b^0
(p^{\prime },s^{\prime };\frac{3}{2})\rangle  \frac{\langle \Xi_b^0 (p^{\prime },s^{\prime
};\frac{3}{2})|\bar{\eta}_{\mu}|0\rangle }{p^{\prime
2}-m^{2}}\nonumber\\
&+&\frac{\langle 0|\eta _{\Xi_b^-}|\Xi_b^- (p,s;\frac{1}{2})\rangle
}{p^{2}-m_{\Xi_b^-}^{2}}\langle \pi^+(q)\Xi_b^-(p,s;\frac{1}{2})|\Xi_b^{0}{}'
(p^{\prime },s^{\prime };\frac{3}{2})\rangle  \frac{\langle \Xi_b^{0}{}'(p^{\prime },s^{\prime
};\frac{3}{2})|\bar{\eta}_{\mu}|0\rangle }{p^{\prime
2}-m'{}^{2}}
 +\ldots ,  \label{eq:SRDecay2b}
\end{eqnarray} 
when we consider the $2S$ state. As is seen, we represent the ground state, $1P$ and $2S$ excited states of spin-$\frac{3}{2}$ baryon, with  the $\Xi_b^0$, $\widetilde{\Xi}_b^0$ and $\Xi_b^0{}'$, respectively as in the previous section. Here,  $\ldots$ corresponds to the contributions coming from the higher states and continuum again. Besides the matrix elements given in Eq.~(\ref{eq:Res2ek}), we need to define additional matrix elements:
\begin{eqnarray}
\langle 0|\eta_{\Xi_b^{-}} |\Xi_b^{-}(p,s;\frac{1}{2})\rangle &=&\lambda_{\Xi_b^-} u(p,s),
\nonumber \\
\langle \pi^+(q)\Xi_b^-(p,s;\frac{1}{2})|\Xi_b^{0}(p^{\prime },s^{\prime };\frac{3}{2})\rangle &=& g_{\Xi_b^{0}\Xi_b^-\pi^+}\bar{u}(p,s)u_{\mu}(p',s')q^{\mu},\nonumber\\
\langle \pi^+(q)\Xi_b^-(p,s;\frac{1}{2})|\widetilde{\Xi}_b^{0}(p^{\prime },s^{\prime };\frac{3}{2})\rangle &=& g_{\widetilde{\Xi}_b^{0}\Xi_b^-\pi^+}\bar{u}(p,s) \gamma_5 u_{\mu}(p',s')q^{\mu},\nonumber\\
\langle \pi^+(q)\Xi_b^-(p,s;\frac{1}{2})|\Xi_b^0{}'(p^{\prime },s^{\prime };\frac{3}{2})\rangle &=& g_{\Xi_b^0{}'\Xi_b^- \pi^+}\bar{u}(p,s)u_{\mu}(p',s')q^{\mu}\label{eq:Res3},
\end{eqnarray}
where $g_{\Xi_b^{0}\Xi_b^-\pi^+}$, $g_{\widetilde{\Xi}_b^{0}\Xi_b^-\pi^+}$ and $g_{\Xi_b^0{}'\Xi_b^- \pi^+}$ are the strong coupling constants: the fundamental parameters that help us estimate the widths of the considering strong decays. In the following, we aim to get sum rules for these strong coupling constants and obtain their numerical values.
Using these new matrix elements inside Eqs.~(\ref{eq:SRDecay2a}) and (\ref{eq:SRDecay2b}) as well as the  summations over spins of $\frac{3}{2}$ particles given in Eq.~(\ref{Rarita}) and summations over spins of $\frac{1}{2}$ states as
\begin{eqnarray}\label{Dirac}
\sum_s  u (p,s)  \bar{u} (p,s) &= &(\!\not\!{p} + m),
\end{eqnarray}
we obtain the hadronic sides as
\begin{eqnarray}
\Pi _{\mu }^{\mathrm{Had}}(p,q)&=&-\frac{g_{\Xi_b^{0} \Xi_b^- \pi^+}\lambda
_{\Xi_b^-}\lambda}{(p^{2}-m_{\Xi_b^-}^{2})(p^{\prime
2}-m{}^{2})}q^{\alpha }(\slashed p+m_{\Xi_b^-})\left( \slashed p'+m\right) T_{\alpha \mu }+\frac{g_{\widetilde{\Xi}_b^{0} \Xi_b^- \pi^+}\lambda
_{\Xi_b^-}\widetilde{\lambda}}{(p^{2}-m_{\Xi_b^-}^{2})(p^{\prime }{}^{2}-%
\widetilde{m}^{2})} \notag \\
&\times & q^{\alpha }(\slashed p+m_{\Xi_b^-})  \gamma _{5}\left( \slashed p'+\widetilde{m}\right) T_{\alpha \mu }\gamma _{5}+\ldots,\label{eq:SRDecayPhys2a}
\end{eqnarray}
\begin{eqnarray}
\Pi _{\mu }^{\mathrm{Had}}(p,q)&=&-\frac{g_{\Xi_b^{0} \Xi_b^- \pi^+}\lambda
_{\Xi_b^-}\lambda}{(p^{2}-m_{\Xi_b^-}^{2})(p^{\prime
2}-m^{2})}q^{\alpha }(\slashed p+m_{\Xi_b^-})\left( \slashed p'+m\right) T_{\alpha \mu }-\frac{g_{\Xi_b^{0}{}' \Xi_b^- \pi^+}\lambda
_{\Xi_b^-}\lambda'}{(p^{2}-m_{\Xi_b^-}^{2})(p^{\prime
2}-m'{}^{2})}\notag\\
&\times & q^{\alpha }(\slashed p+m_{\Xi_b^-})\left( \slashed p'+m'\right) T_{\alpha \mu }+\ldots\label{eq:SRDecayPhys2b},
\end{eqnarray}%
where the expression $T_{\alpha\mu}$ is given as
\begin{eqnarray}
&&T_{\alpha \mu }(p)=g_{\alpha \mu }-\frac{1}{3}\gamma _{\alpha }\gamma
_{\mu }-\frac{2}{3m^{2}}p_{\alpha }p_{\mu }+\frac{1}{3m}\left[ p_{\alpha }\gamma _{\mu }-p_{\mu
}\gamma _{\alpha }\right] .
\end{eqnarray}%
With an application of double Borel transformation with respect to $p'{}^2$ and $p^2$ with corresponding Borel parameters $M_1^2$ and $M_2^2$, the results of the hadronic sides turn into
\begin{eqnarray}
{\cal B} \Pi _{\mu }^{\mathrm{Had}}(p,q)&=&-g_{\Xi_b^{0} \Xi_b^- \pi^+}\lambda _{\Xi_b^-}\lambda e^{-\frac{m%
^{2}}{M_{1}^{2}}}e^{-\frac{m_{\Xi_b^-}^{2}}{M_{2}^{2}}}q^{\alpha }  (\slashed p+m_{\Xi_b^-})\left( \slashed p'+%
m\right) T_{\alpha \mu }\notag \\
&+&g_{\widetilde{\Xi}_b^{0} \Xi_b^- \pi^+}\lambda _{\Xi_b^-}\widetilde{\lambda}  
 e^{-\frac{\widetilde{m}{}^2}{M_{1}^{2}}}e^{-\frac{m_{\Xi_b^-}^{2}}{M_{2}^{2}}}q^{\alpha
}(\slashed p+m_{\Xi_b^-})\gamma _{5}\left( \slashed p'+\widetilde{m} \right)
T_{\alpha \mu }\gamma _{5}. \label{eq:CFunc3/2a}
\end{eqnarray}
\begin{eqnarray}
{\cal B} \Pi _{\mu }^{\mathrm{Had}}(p,q)&=&-g_{\Xi_b^{0} \Xi_b^- \pi^+}\lambda _{\Xi_b^-}\lambda e^{-\frac{m%
^{2}}{M_{1}^{2}}}e^{-\frac{m_{\Xi_b^-}^{2}}{M_{2}^{2}}}q^{\alpha }  (\slashed p+m_{\Xi_b^-})\left( \slashed p'+%
m\right) T_{\alpha \mu }\notag \\
&-&g_{\Xi_b^{0}{}' \Xi_b^- \pi^+}\lambda _{\Xi_b^-}\lambda' e^{-\frac{m'%
^{2}}{M_{1}^{2}}}e^{-\frac{m_{\Xi_b^-}^{2}}{M_{2}^{2}}}q^{\alpha }  (\slashed p+m_{\Xi_b^-})\left( \slashed p'+%
m'\right) T_{\alpha \mu },  \label{eq:CFunc3/2b}
\end{eqnarray}
where, to represent the Borel transformed results of the correlation function, we used ${\cal B} \Pi _{\mu }^{\mathrm{Had}}(p,q)$. Among various  Lorentz structures we consider $\slashed p q_{\mu}$ and $\slashed q q_{\mu}$  to evaluate the corresponding strong coupling constants. We should also  note here that, before selection of the structures, we order the dirac matrices as  $\slashed q \slashed p\gamma_{\mu}$ and remove the spin-1/2 pollution (in this case only in the initial states), which couple to the spin-3/2 current, from a similar way explained in the previous section.  

To calculate the  QCD side of the calculations, we use the related interpolating currents explicitly inside the correlator and make the possible contractions between quark fields via Wick's theorem. As a result the expressions are obtained in terms of the heavy and light quark propagators and the matrix elements of the remaining quark fields after contractions, such as $\langle \pi^+(q)|\bar{q}(x)\Gamma q(y)|0\rangle$ or $\langle \pi^+(q)|\bar{q}(x)\Gamma G_{\mu\nu} q(y)|0\rangle$.  The $\Gamma$ and $G_{\mu\nu}$ present in these matrix elements are the full set of Dirac matrices and gluon field strength tensor, respectively. These matrix elements are described in terms of $\pi$-meson distribution amplitudes (DAs), and their usage in the calculations leads to the nonperturbative contributions. For the explicit expressions of these matrix elements see the  Refs.~\cite{Ball:2006wn,Belyaev:1994zk,Ball:2004ye,Ball:2004hn}. The results obtained from this side are quite long mathematical expressions, and therefore we will not write them here. The coefficients of the same structures from both sides are again matched, and Borel transformation and continuum subtraction are applied to the QCD side. After these procedures, the QCD sum rules giving the corresponding  strong coupling constants of the considered decays are attained as  
\begin{eqnarray}
{\cal B}\Pi_{1}^{\mathrm{QCD,tr}}&=&-g_{\Xi_b^{0}\Xi_b^- \pi^+}  \lambda_{\Xi_{b}^- } \lambda  \frac{[(m+m_{\Xi_b^-})(2m^2+m_{\Xi_b^-}^2)-m_{\pi^+}^2(2m+m_{\Xi_b^-})]}{3m^2}e^{-\frac{m^2}{M_1^2}}e^{-\frac{m_{\Xi_b^-}^2}{M_2^2}}
\nonumber\\&+&
g_{\widetilde{\Xi}b^{0}\Xi_b^- \pi^+}\lambda{\Xi_{b}^- } \widetilde{\lambda}  \frac{[(\widetilde{m}-m_{\Xi_b^-})(2\widetilde{m}^2+m_{\Xi_{b}^-}^2)-m_{\pi^+}^2(2\widetilde{m}-m_{\Xi_b^-})]}{3\widetilde{m}^2}e^{-\frac{\widetilde{m}^{2}}{M_{1}^{2}}}e^{-\frac{m_{\Xi_b^-}^2}{M_2^2}},\nonumber\\
{\cal B}\Pi_{2}^{\mathrm{QCD,tr}}&=&-g_{\Xi_b^{0}\Xi_b^- \pi^+}
\lambda_{\Xi_{b}^- } \lambda \frac{[m^2+m_{\Xi_{b}^-}^2-m
m_{\Xi_{b}^-}-m_{\pi^+}^2]m_{\Xi_{b}^-}}{3m^2}e^{-\frac{m^2}{M_1^2}}e^{-\frac{m^2_{\Xi_b^-}}{M_2^2}}\nonumber\\&-&g_{\widetilde{\Xi}b^{0}\Xi_b^-
\pi^+}\lambda{\Xi_{b}^- } \widetilde{\lambda}
\frac{[\widetilde{m}^2+m_{\Xi_{b}^-}^2+\widetilde{m}m_{\Xi_{b}^-}-m_{\pi^+}^2]m_{\Xi_{b}^-}}{3\widetilde{m}^2}
e^{-\frac{\widetilde{m}^{2}}{M_{1}^{2}}}e^{-\frac{m_{\Xi_b^-}^2}{M_2^2}}\label{eq:couplingpair3/2a},
\end{eqnarray}
for the ground $ +1 P $ and 
\begin{eqnarray}
{\cal B}\Pi_{1}^{\mathrm{QCD,tr}}&=&-g_{\Xi_b^{0}\Xi_b^-\pi^+}  \lambda_{\Xi_{b}^- }\lambda \frac{[(m+m_{\Xi_{b}^-})(2m^2+m_{\Xi_b^0}^2)-m_{\pi^+}^2(2m+m_{\Xi_b^-})]}{3m^2} e^{-\frac{m^2}{M_1^2}}e^{-\frac{m_{\Xi_b^-}^2}{M_2^2}}
\nonumber\\&-& 
g_{\Xi_b^{0}{}'\Xi_b^- \pi^+} \lambda_{\Xi_{b}^- }\lambda'\frac{[(m'+m_{\Xi_b^-})(2m'^2+m_{\Xi_{b}^-}^2)-m_{\pi^+}^2(2m'+m_{\Xi_b^-})]}{3m'^2} e^{-\frac{m'{}^{2}}{M_{1}^{2}}}e^{-\frac{m_{\Xi_b^-}^2}{M_2^2}},\nonumber\\
{\cal B}\Pi_{2}^{\mathrm{QCD,tr}}&=&-g_{\Xi_b^{0}\Xi_b^- \pi^+}
\lambda_{\Xi_{b}^- } \lambda \frac{[m^2+m_{\Xi_{b}^-}^2-m
m_{\Xi_{b}^-}-m_{\pi^+}^2]m_{\Xi_{b}^-}}{3m^2}e^{-\frac{m^2}{M_1^2}}e^{-\frac{m_{\Xi_b^-}^2}{M_2^2}}\nonumber\\&-&g_{\Xi_b^{0}{}'\Xi_b^-
\pi^+}\lambda_{\Xi_{b}^- } \lambda'
\frac{[m'{}^2+m_{\Xi_{b}^-}^2-m'
m_{\Xi_{b}^-}-m_{\pi^+}^2]m_{\Xi_{b}^-}}{3m'{}^2}
e^{-\frac{m'{}^{2}}{M_{1}^{2}}}e^{-\frac{m_{\Xi_b^-}^2}{M_2^2}},\label{eq:couplingpair3/2b}
\end{eqnarray}
for the ground $ +2 S $ case.
In Eqs.~(\ref{eq:couplingpair3/2a}) and (\ref{eq:couplingpair3/2b}), the ${\cal B}\Pi_{1}^{\mathrm{QCD,tr}}$ and ${\cal B}\Pi_{2}^{\mathrm{QCD,tr}}$ are used to represent the results obtained from QCD side of the transition considered in this section  after the Borel transformation.  They are the coefficients of the Lorentz structures  $\slashed p q_{\mu}$ and $\slashed q q_{\mu}$, respectively.

\begin{figure}[h!]
\begin{center}
\includegraphics[totalheight=5cm,width=7cm]{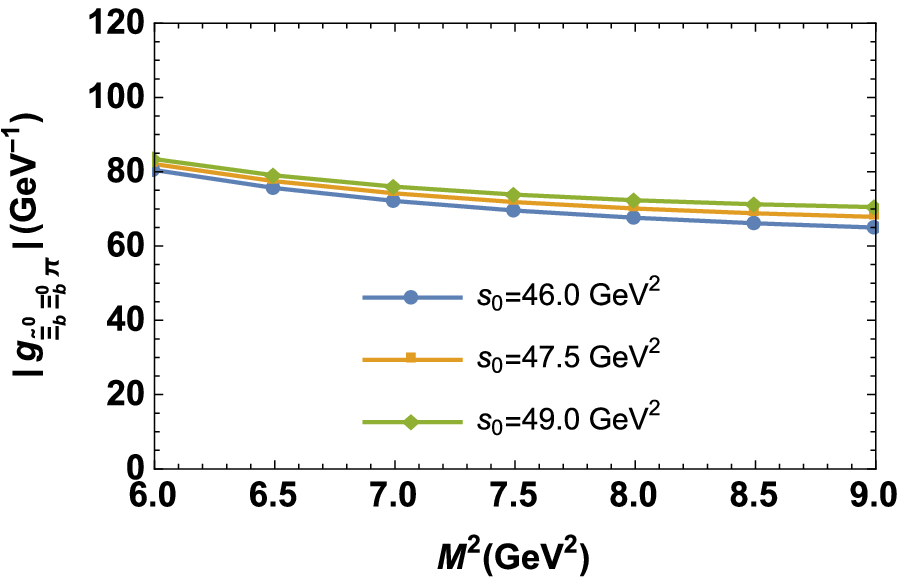}
\includegraphics[totalheight=5cm,width=7cm]{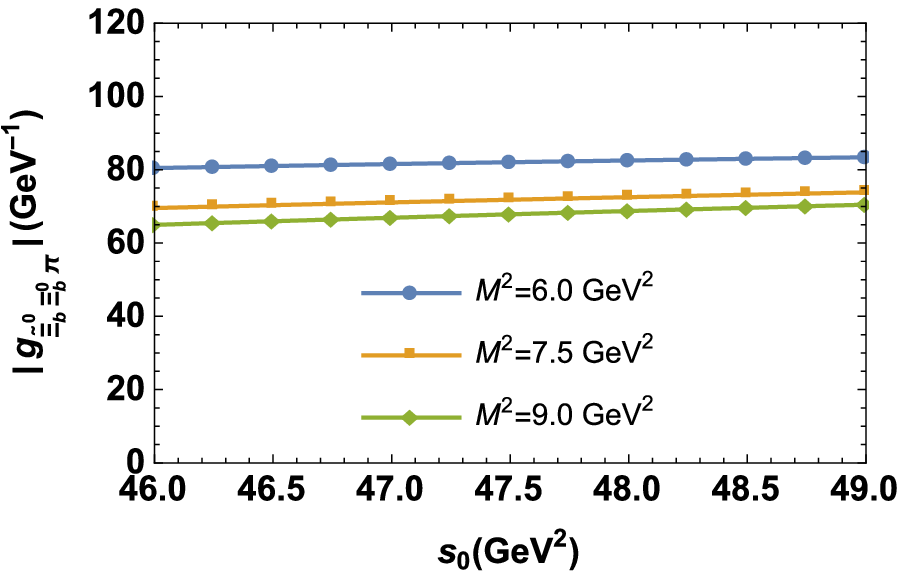}
\end{center}
\caption{\textbf{Left:} The coupling constant $g_{\widetilde{\Xi}_b^0\Xi_b^- \pi^+}$ obtained for the $ 1P $ excitation of $\Xi_b(5945)^{0}$ vs Borel parameter $M^2$.
\textbf{Right:} The coupling constant $g_{\widetilde{\Xi}_b^0 \Xi_b^- \pi^+}$ obtained for the $ 1P $ excitation of $\Xi_b(5945)^{0}$ vs threshold parameter $s_0$. }
\label{gr:gXib}
\end{figure}

To analyze the obtained results we adopt the input parameters given in Table~\ref{tab:Param} and due to the close masses of  the initial and final states we use $M_1^2=M_2^2$ and $M^2=\frac{M_1^2 M_2^2}{M_1^2+M_2^2}$ from which we get $M_1^2=M_2^2=2M^2$. For the auxiliary parameters, we exploit the ones that we used in the mass calculations. Our analyses show that the considered first two resonances contributions (FTRC) in each case constitute the main part of the total contribution, satisfying the requirement of the method. In the case of $\slashed q q_{\mu}$, for instance, the FTRC is obtained to be $ 83\% $. The remaining $ 17\% $ belongs to all the higher states and continuum.  In Fig.~\ref{gr:gXib} the graphs of $g_{\tilde{\Xi}_b^0 \Xi_b^- \pi^+}$ as functions of $M^2$ and $s_0$ are presented to illustrate its variation with respect to these auxiliary parameters. From these figures, we see that the strong coupling constants weakly depend on the $M^2$ and $s_0$ in their working intervals.  
The coupling constants obtained for each case, that is, for the decays of the $1P$ and $2S$ initial states, are used to obtain the corresponding decay widths. To this end, the following decay width formulas are applied
\begin{eqnarray}
\Gamma (\widetilde{\Xi}_b^{0} &\rightarrow &\Xi_b^- \pi^+)=\frac{%
g_{\widetilde{\Xi}_b^{0}\Xi_b^- \pi^+}^{2}}{24\pi \widetilde{m}^{2}}\left[(\widetilde{m}%
-m_{\Xi_b^- })^{2}-m_{\pi^+}^{2}\right]  f^3(\widetilde{m},m_{\Xi_b^-},m_{\pi^+}),
\end{eqnarray}%
for the decay of $1P$ state and
\begin{eqnarray}
\Gamma (\Xi_b^{ 0}{}' &\rightarrow &\Xi_b^- \pi^+)=\frac{%
g_{\Xi_b^{0}{}'\Xi_b^- \pi^+}^{2}}{24\pi m'{}^2}\left[ (m'
+m_{\Xi_b^-})^{2}-m_{\pi^+}^{2}\right]  f^3(m',m_{\Xi_b^-},m_{\pi^+}),
\end{eqnarray}
for the decay of $2S$ state, where $f(x,y,z)$ is 
\begin{equation*}
f(x,y,z)=\frac{1}{2x}\sqrt{%
x^{4}+y^{4}+z^{4}-2x^{2}y^{2}-2x^{2}z^{2}-2y^{2}z^{2}}.
\end{equation*}
The numerical values of the strong coupling constant and partial decay width for the decay of $1P$ state are obtained as
\begin{eqnarray}
g_{\widetilde{\Xi}_b^{0}\Xi_b^- \pi^+}=73.82\pm 4.99~\mathrm{GeV}^{-1}, ~~~~~~~~~ \Gamma(\widetilde{\Xi}_b^{0}, \rightarrow \Xi_b^- \pi^+)=18.31 \pm 2.56~\mathrm{MeV},
\end{eqnarray}
and those of $2S$ state are
\begin{eqnarray}
g_{\Xi_b^{0}{}' \Xi_b^- \pi^+}=52.75\pm 3.71~\mathrm{GeV}^{-1}, ~~~~~~~~~ \Gamma(\Xi_b^{0}{}' \rightarrow  \Xi_b^- \pi^+)=8.24 \pm 1.19~\mathrm{MeV}.
\end{eqnarray}

From these results, among the two possible scenarios, the partial  width for $1P$ excitation case is seen to be consistent with the experimental observation, $\Gamma(\Xi_b(6227)^0)=18.6^{+5.0}_{-4.1}\pm 1.4~\mathrm{MeV}$~\cite{Aaij:2020fxj}, supporting  the possibility that the observed state's quantum numbers  would be $J^P=\frac{3}{2}^-$.   This assigned is made based on the comparison of the obtained results for the partial widths of the decays $\widetilde{\Xi}_b^{0}, \rightarrow \Xi_b^- \pi^+  $ and $\Xi_b^{0}{}' \rightarrow  \Xi_b^- \pi^+  $ considering them as dominant decays at different scenarios. If other decay modes are observed, the results of widths should be taken into account for exact determination of the quantum numbers for $\Xi_b(6227)^0$ state.

\section{Summary and conclusion}\label{Sum} 

This work is performed to shed light on the quantum numbers of the newly observed $\Xi_b(6227)^0$ state. Recently, the LHCb announced the observation of a new excited resonance, $\Xi_b(6227)^0$, and an improvement for the mass and width measurements for the $\Xi_b(6227)^-$ state~\cite{Aaij:2020fxj}. Following the first observation of the $\Xi_b(6227)^-$~\cite{Aaij:2018yqz}, the masses and decay widths for the various possible scenarios with different possible quantum numbers were considered in Ref.~\cite{Aliev:2018lcs}, and the obtained results were in favor of its quantum numbers being $J^P=\frac{3}{2}^-$. In this study, considering our previous investigation of $\Xi_b(6227)^-$ state and the experimental information that the new state $\Xi_b(6227)^0$ is the isospin partner of  $\Xi_b(6227)^-$, we assigned the possible spin quantum number for the newly observed $\Xi_b(6227)^0$ state as $J=\frac{3}{2}$. We considered the two possibilities that  this state can be either $1P$ or $2S$ excitation of the ground state $\Xi_b(5945)^0$ baryon. With these considerations, we made  QCD sum rules analyses for the corresponding masses and partial widths of the decays of these excited states. The results of masses and partial decay widths were obtained as $\tilde{m}=6225.47 \pm 106.29~\mathrm{MeV}$ and $\Gamma(\widetilde{\Xi}_b^{0}, \rightarrow \Xi_b^- \pi^+)=18.13 \pm 2.35~\mathrm{MeV}$ for $1P$ state and as $m'=6225.47 \pm 106.29~\mathrm{MeV}$ and $\Gamma(\Xi_b^{0}{}' \rightarrow  \Xi_b^- \pi^+)=5.43 \pm 0.70~\mathrm{MeV}$ for $2S$ state. Though both masses are consistent with the experimental observation, the partial width obtained for $\widetilde{\Xi}_b^{0}$ case, which is in nice agreement with the experimental value  $\Gamma(\Xi_b(6227)^0)=18.6^{+5.0}_{-4.1}\pm 1.4~\mathrm{MeV}$~\cite{Aaij:2020fxj}, supports the possibility to assign the quantum numbers of  $\Xi_b(6227)^0$ state to be $J^P=\frac{3}{2}^-$. Note that this assignment   is subject to change: We made it by comparison of the partial width of the decay mode, and possibly the dominant one, that the  $\Xi_b(6227)^0$ state has been observed. If another sizable decay mode is observed, the corresponding width should be added to the one that is considered in the present study.

\section*{ACKNOWLEDGEMENTS}
K. Azizi is thankful to Iran Science Elites Federation (Saramadan)
for the partial  financial support provided under the grant number ISEF/M/99171.

\section*{Appendix: The QCD side of the mass sum rules}

In this appendix, as an example,  we present the explicit expression for the function $  \Pi_1^{\mathrm{QCD}} $ (see Eq. (\ref{Eq:sumrule2})), obtained from the mass sum rules in section II. It is obtained as
\begin{eqnarray}
 \Pi_1^{\mathrm{QCD}}=\int_{(m_b+m_s)^2}^{s_0} e^{-\frac{s}{M^2}}\rho_1(s) ds + e^{-\frac{m_b^2}{M^2}}\Gamma_1+ e^{-\frac{s_0}{M^2}}\tilde{\Gamma}_1,
\end{eqnarray}
where,
\begin{eqnarray}
\rho_1(s)& = & 
-\frac{m_b^3 (m_s + m_u) }{
  192 \pi^4}\Big[6 \psi_{10} - 3 \psi_{20}  + 3 \psi_{31} -\psi_{32} + 6 \ln\big(\frac{m_b^2}{s}\big)\Big] - \frac{
 m_b^4 }{3840 \pi^4}\Big[-60 \psi_{10} + 30 \psi_{20} - 20 \psi_{30} + 100 \psi_{41} \nonumber\\ &+& 31 \psi_{42} + 4 \psi_{43} + 
    60 \psi_{00} \ln\big(\frac{s}{m_b^2}\big)\Big]+\langle\bar{u}u\rangle\Big[-\frac{ m_0^2}{48 m_b \pi^2}\psi_{02} +\frac{m_b}{24 \pi^2} (2 \psi_{10} - \psi_{11} -\psi_{12} - 2 \psi_{21})  +\frac{m_0^2 m_u}{96 m_b^2 \pi^2} (\psi_{02} - 4 \psi_{03}) \nonumber\\ &+&  \frac{1}{48\pi^2} \big[4 (-\psi_{00} + \psi_{03}) m_u + (\psi_{11} + \psi_{12}) (2 m_s + 3 m_u)\big]\Big]
-\langle \bar{s}s\rangle \Big[ -\frac{ m_0^2}{48 m_b \pi^2}\psi_{02} + \frac{m_b}{24 \pi^2}(2 \psi_{10} - \psi_{11} - \psi_{12} - 2 \psi_{21}) \nonumber\\
& +&
\frac{m_0^2 m_s}{96 m_b^2 \pi^2} (\psi_{02} - 4 \psi_{03} ) + \frac{1}{48 \pi^2}\big[-4 \psi_{00} m_s + 
  4 \psi_{03} m_s + (\psi_{11} + \psi_{12}) (3 m_s + 2 m_u)\big]\Big]
 +
 \frac{\langle g^2 G^2\rangle }{1152 m_b \pi^4} \Big[\psi_{03} m_b + 4 \psi_{10} m_b \nonumber\\
 &-& 4 \psi_{21} m_b - 3 \psi_{10} m_s + 2 \psi_{11} m_s + 2 \psi_{12} m_s + 3 \psi_{21} m_s - 3 \psi_{10} m_u + 2 \psi_{11} m_u + 2 \psi_{12} m_u + 3 \psi_{21} m_u \nonumber\\
 &+& (3 \psi_{01} - \psi_{02}) (m_s + m_u) - \psi_{00} (m_b + 3 \gamma_E (m_s + m_u)) + 3 \psi_{02} (m_s + m_u) \big[\ln\big(\frac{\Lambda^2}{s}\big) + \ln\big(\frac{(s-m_b^2)}{\Lambda^2}\big) \nonumber\\
 &+& \ln\big(\frac{s (s-m_b^2 )}{\Lambda^2 m_b^2}\big)\big]\Big]
 +\frac{\langle g^2G^2\rangle }{1728 m_b^4 M^{10} \pi^2}\Big[ 2 M^{10} \big[-3 (2 \psi_{01} - 2 \psi_{12} - 5 \psi_{13} - 11 \psi_{14} - 2 \psi_{23}) (m_u \langle \bar{s}s\rangle + m_s \langle \bar{u}u\rangle)\nonumber\\
 & +& \psi_{03} (4 m_s \langle \bar{s}s\rangle - 6 m_u \langle \bar{s}s\rangle - 6 m_s \langle \bar{u}u\rangle + 4 m_u \langle \bar{u}u\rangle) + 3 \psi_{02} (m_b (\langle \bar{s}s\rangle + \langle \bar{u}u\rangle) - 9 (m_u \langle \bar{s}s\rangle + m_s \langle \bar{u}u\rangle))\big]\nonumber\\
&+&  3 (m_u \langle \bar{s}s\rangle + 
  m_s \langle \bar{u}u\rangle) \big[\psi_{00} \big(m_0^2 m_b^8 - 2 M^4 (2 m_b^6 + 3 m_b^4 M^2 + 2 m_b^2 M^4 + 6 M^6)\big) \ln\big(\frac{ (s-m_b^2)}{\Lambda^2}\big) \nonumber\\
  &+& 12 \psi_{04} M^{10} \ln\big(\frac{s (s-m_b^2)}{\Lambda^2 m_b^2}\big)\big]\Big] 
-\frac{\langle g^2G^2 \rangle ^2}{41472 M^{16} \pi^2}\Big[ (m_u  \langle \bar{s}s\rangle + 
    m_s  \langle \bar{u}u\rangle)\psi_{00} m_b^2 \big(2 M^4 (-2 m_b^2 + 5 M^2) \nonumber\\
    &+& 
    m_0^2 (m_b^4 - 7 m_b^2 M^2 + 9 M^4)\big)   \ln\big(\frac{ (s-m_b^2)}{\Lambda^2}\big)\Big] ,\label{eq:couplingpair3/2bbb}
\end{eqnarray}
\begin{eqnarray}
\Gamma_1&=&\frac{m_0^2}{288 \pi^2} (7 m_s \langle \bar{s}s\rangle - 6 m_u \langle \bar{s}s\rangle - 6 m_s \langle \bar{u}u\rangle + 7 m_u \langle \bar{u}u\rangle)+\frac{\langle \bar{u}u\rangle \langle \bar{s}s\rangle}{216 M^6}\Big[12 (m_b (m_s + m_u) - 2 M^2) M^4 \nonumber\\
&+& 
m_0^2 (-5 m_b^3 (m_s + m_u) + 12 m_b^2 M^2 + 12 M^4)\Big] +\frac{\langle g^2G^2\rangle}{384 m_b \pi^4} \gamma_E (m_s + m_u) M^2
-\frac{\langle g^2G^2\rangle ^2}{82944 m_b^3 M^4 \pi^4}\Big[2 m_b^4 (m_s + m_u) \nonumber\\
&+& m_b^2 \big(m_b + (-1 + 3 \gamma_E) (m_s + m_u)\big) M^2 + 
 2 \big(m_b - 3 (-1 +\gamma_E) (m_s + m_u)\big) M^4 +  3 (m_s + m_u)\big(m_b^4 - m_b^2 M^2 \nonumber\\
 &-& M^4\big) \ln\big(\frac{ \Lambda^2}{m_b^2}\big)\Big]
 +\frac{\langle g^2 G^2\rangle}{62208 m_b^2 M^{12} \pi^2}\Big[-72 M^8 \langle \bar{s}s\rangle \big[2 m_b^4 m_u + m_b^2 M^2 (-2 m_s + 3 (2 + \gamma_E) m_u)  + (m_b + 2 m_s + 6 m_u \nonumber\\
 &+& 12 \gamma_E m_u) M^4\big]- 24 M^4 \langle \bar{u}u\rangle \big[3 M^4 \big(2 m_b^4 m_s +m_b^2 M^2 (3 (2 +\gamma_E ) m_s - 2 m_u)  +  M^4 (m_b + 6 m_s +12 \gamma_E m_s + 2 m_u)\big) \nonumber\\
 &+& 2 m_b^3  \pi^2 \langle \bar{s}s\rangle \big(m_b^2 (m_s + m_u) - 2 m_b M^2 -3 (m_s + m_u) M^2\big)\big] + m_0^2  \big[9 M^4  \langle \bar{s}s\rangle \big[4 m_b^6 m_u + m_b^4 (-3 m_s + 20 m_u) M^2\nonumber\\
 &+& 2 m_b^2 (m_b + m_s + 9 m_u) M^4 - 4 (m_b - 3 m_u) M^6\big]  + \langle \bar{u}u\rangle \big[9 M^4 \big(4 m_b^6 m_s + m_b^4 (20 m_s - 3 m_u) M^2 \nonumber\\
 &+& 2 m_b^2 (m_b + 9 m_s + m_u) M^4 -4 (m_b - 3 m_s) M^6\big) +4 m_b^3 \pi^2  \langle \bar{s}s\rangle \big(5 m_b^4 (m_s + m_u) -6 m_b^2 (2 m_b + 5 (m_s + m_u)) M^2 \nonumber\\
&+& 6 (4 m_b + 5 (m_s + m_u)) M^4\big)  \big] \big] -54 M^4(m_u  \langle \bar{s}s\rangle+ m_s  \langle \bar{u}u\rangle) (-m_0^2 m_b^6 + 4 M^4 (m_b^4 + 2 m_b^2 M^2 + 2 M^4))  \ln\big(\frac{ \Lambda^2}{m_b^2}\big)\Big]\nonumber\\
&-&\frac{ \langle g^2 G^2\rangle ^2}{248832 m_b^2 M^{14} \pi^2}(m_u \langle \bar{s}s\rangle + m_s \langle \bar{u}u\rangle) \Big[4 M^4 \big(-2 m_b^6 + (3 \gamma_E -8  ) m_b^4 M^2 +9 m_b^2 M^4 + 3 M^6\big) \nonumber\\ 
&+&m_0^2 \big(2 m_b^8 - 6 ( \gamma_E - 1) m_b^6 M^2 +3 (6 \gamma_E -19 ) m_b^4 M^4 + 24 m_b^2 M^6 + 6 M^8\big) +3 m_b^4 \big[m_0^2 (m_b^2 - 6 M^2) (m_b^2 - 2 M^2) \nonumber\\
&+& 4 M^4 (-m_b^2 + 3 M^2)\big] \ln\big(\frac{ \Lambda^2}{m_b^2}\big)\Big],
\end{eqnarray}
and
\begin{eqnarray}
\tilde{\Gamma}_1&=&-\frac{\langle g^2 G^2\rangle}{384 m_b \pi^4}\Big[\gamma_E (m_s + m_u) M^2\Big]-\frac{\langle g^2G^2\rangle ^2 (m_s + m_u)}{27648 m_b M^4 \pi^4 (m_b^2 - s_0)^2}\Big[ M^2 (-2 m_b^4 - 3 M^2 s_0 + m_b^2 (M^2 + 2 s_0)) \nonumber\\
&+& (2 m_b^2 - 3 M^2) (m_b^2 - s_0)^2 \ln\big(\frac{(s_0-m_b^2)}{\Lambda^2}\Big)\Big]+\frac{\langle g^2 G^2\rangle}{576 M^8 \pi^2}(m_u \langle \bar{s}s\rangle + m_s \langle \bar{u} u\rangle)\Big[ \frac{1}{(m_b^2 - s_0)^4}\big(-6 m_0^2 m_b^4 M^8 \nonumber\\
&+& (m_0^2 m_b^4 M^4 - 4 m_b^2 M^8) (m_b^2 - s_0)^2 + 2 m_o^2 m_b^4 M^6 (-m_b^2 + s_0) + (m_0^2 m_b^4 M^2 - 2 M^6 (2 m_b^2 + 3 M^2)) (-m_b^2 + s_0)^3 \big)\nonumber\\
&+& (m_0^2 m_b^4 - 2 M^4 (2 m_b^2 + 3 M^2))   \ln\big(\frac{(s_0-m_b^2)}{\Lambda^2}\Big) \Big]
+\frac{\langle g^2 G^2\rangle ^2}{41472 M^{14} \pi^2 (m_b^2 - s_0)^7} m_b^2  (m_u \langle \bar{s}s\rangle + m_s\langle \bar{u}u \rangle) \nonumber\\
&\times &
 \Big[ -2 M^6 (m_b^2 - s_0)^2 \big[2 m_b^{10} - m_b^8 (3 M^2 + 8 s_0) + m_b^6 (-M^4 + 14 M^2 s_0 + 12 s_0^2) +m_b^4 (2 M^6 + 7 M^4 s_0 - 24 M^2 s_0^2 - 8 s_0^3) \nonumber\\
 & +& 5 M^2 s_0 (6 M^6 - 2 M^4 s_0 + M^2 s_0^2 - s_0^3) + m_b^2 (18 M^8 + 8 M^6 s_0 - 11 M^4 s_0^2 + 18 M^2 s_0^3 +2 s_0^4)\big] + 
 m_0^2 M^2 \big[m_b^{16} \nonumber\\
 &-& 6 m_b^{14} (M^2 + s_0) + m_b^{12} (4 M^4 + 37 M^2 s_0 + 15 s_0^2) + m_b^{10} (M^6 - 27 M^4 s_0 - 95 M^2 s_0^2 - 20 s_0^3) +m_b^8 s_0 (-7 M^6 \nonumber\\
 &+& 77 M^4 s_0 + 130 M^2 s_0^2 + 15 s_0^3) + 9 M^4 s_0^2 (24 M^8 - 6 M^6 s_0 + 2 M^4 s_0^2 - M^2 s_0^3 + 
s_0^4) + m_b^2 M^2 s_0 (408 M^{10} - 6 M^8 s_0 \nonumber\\
&-& 30 M^6 s_0^2 +31 M^4 s_0^3 - 47 M^2 s_0^4 - 7 s_0^5) +2 m_b^6 (3 M^{10} + 3 M^8 s_0 + 12 M^6 s_0^2 - 59 M^4 s_0^3 -50 M^2 s_0^4 - 3 s_0^5) + m_b^4 (96 M^{12} \nonumber\\
&+& 54 M^{10} s_0 + 6 M^8 s_0^2 - 40 M^6 s_0^3 +102 M^4 s_0^4 + 41 M^2 s_0^5 + s_0^6)\big] - \big[2 M^4 (-2 m_b^2 + 5 M^2) + m_0^2 (m_b^4 - 7 m_b^2 M^2 \nonumber\\
&+& 9 M^4)\big] (m_b^2 - s_0)^7  \ln\big(\frac{(s_0-m_b^2)}{\Lambda^2}\Big)  \Big],
\end{eqnarray}
with
\begin{eqnarray}
\psi_{nm}=\frac{(s-m_b^2)^n}{s^m(m_b^2)^{n-m}}.
\end{eqnarray}




\begin{thebibliography}{99}

 \bibitem{Aaij:2012da} 
  R.~Aaij {\it et al.} [LHCb Collaboration],
  \href{https://journals.aps.org/prl/pdf/10.1103/PhysRevLett.109.172003}{Phys.\ Rev.\ Lett.\  {\bf 109}, 172003 (2012)}
 \href{https://arxiv.org/abs/1205.3452}{[arXiv:1205.3452 [hep-ex]]}.
 
  
  \bibitem{Chatrchyan:2012ni} 
  S.~Chatrchyan {\it et al.} [CMS Collaboration],
\href{https://journals.aps.org/prl/abstract/10.1103/PhysRevLett.108.252002}{  Phys.\ Rev.\ Lett.\  {\bf 108}, 252002 (2012)}
 \href{https://arxiv.org/abs/1204.5955}{ [arXiv:1204.5955 [hep-ex]]}.
  
  \bibitem{Aaij:2014yka} 
  R.~Aaij {\it et al.} [LHCb Collaboration],
 \href{https://journals.aps.org/prl/abstract/10.1103/PhysRevLett.114.062004}{ Phys.\ Rev.\ Lett.\  {\bf 114}, 062004 (2015)}
\href{https://arxiv.org/abs/1411.4849}{[arXiv:1411.4849 [hep-ex]]}.


\bibitem{Aaij:2015yoy}  
 R.~Aaij {\it et al.} [LHCb Collaboration],
\href{https://journals.aps.org/prl/abstract/10.1103/PhysRevLett.115.241801}{  Phys.\ Rev.\ Lett.\  {\bf 115}, no. 24, 241801 (2015)}
 \href{https://arxiv.org/abs/1510.03829}{ [arXiv:1510.03829 [hep-ex]]}.
  
  \bibitem{Aaij:2018yqz} 
  R.~Aaij {\it et al.} [LHCb Collaboration],
\href{https://journals.aps.org/prl/abstract/10.1103/PhysRevLett.121.072002}{  Phys.\ Rev.\ Lett.\  {\bf 121}, no. 7, 072002 (2018)}
\href{https://arxiv.org/abs/1805.09418}{  [arXiv:1805.09418 [hep-ex]]}.
  
 
\bibitem{Aaij:2020cex}
R.~Aaij \textit{et al.} [LHCb],
\href{https://journals.aps.org/prl/abstract/10.1103/PhysRevLett.124.082002}{Phys. Rev. Lett. \textbf{124}, no.8, 082002 (2020)}
\href{https://arxiv.org/abs/2001.00851}{[arXiv:2001.00851 [hep-ex]]}.

\bibitem{Aaij:2018tnn}
R.~Aaij \textit{et al.} [LHCb],
\href{https://journals.aps.org/prl/abstract/10.1103/PhysRevLett.122.012001}{Phys. Rev. Lett. \textbf{122}, no.1, 012001 (2019)}
\href{https://arxiv.org/abs/1809.07752}{[arXiv:1809.07752 [hep-ex]]}.


\bibitem{Aaij:2019amv}
R.~Aaij \textit{et al.} [LHCb],
\href{https://journals.aps.org/prl/abstract/10.1103/PhysRevLett.123.152001}{Phys. Rev. Lett. \textbf{123}, no.15, 152001 (2019)}
\href{https://arxiv.org/abs/1907.13598}{[arXiv:1907.13598 [hep-ex]]}.


\bibitem{Capstick:1986bm}
S.~Capstick and N.~Isgur,
\href{https://aip.scitation.org/doi/abs/10.1063/1.35361}{AIP Conf. Proc. \textbf{132}, 267-271 (1985)}.



\bibitem{Ebert:2007nw}
D.~Ebert, R.~N.~Faustov and V.~O.~Galkin,
\href{https://www.sciencedirect.com/science/article/pii/S0370269307014402?via%3Dihub}{Phys. Lett. B \textbf{659}, 612-620 (2008)}
\href{https://arxiv.org/abs/0705.2957}{[arXiv:0705.2957 [hep-ph]]}.


\bibitem{Garcilazo:2007eh}
H.~Garcilazo, J.~Vijande and A.~Valcarce,
\href{https://iopscience.iop.org/article/10.1088/0954-3899/34/5/014}{J. Phys. G \textbf{34}, 961-976 (2007)}
\href{https://arxiv.org/abs/hep-ph/0703257}{[arXiv:hep-ph/0703257 [hep-ph]]}.

\bibitem{Roberts:2007ni}
W.~Roberts and M.~Pervin,
\href{https://www.worldscientific.com/doi/abs/10.1142/S0217751X08041219}{Int. J. Mod. Phys. A \textbf{23}, 2817-2860 (2008)}
\href{https://arxiv.org/abs/0711.2492}{[arXiv:0711.2492 [nucl-th]]}.


\bibitem{Valcarce:2008dr}
A.~Valcarce, H.~Garcilazo and J.~Vijande,
\href{https://link.springer.com/article/10.1140/epja/i2008-10616-4}{Eur. Phys. J. A \textbf{37}, 217-225 (2008)}
\href{https://arxiv.org/abs/0807.2973}{[arXiv:0807.2973 [hep-ph]]}.


\bibitem{Ebert:2011kk}
D.~Ebert, R.~N.~Faustov and V.~O.~Galkin,
\href{https://journals.aps.org/prd/abstract/10.1103/PhysRevD.84.014025}{Phys. Rev. D \textbf{84}, 014025 (2011)}
\href{https://arxiv.org/abs/1105.0583}{[arXiv:1105.0583 [hep-ph]]}.


\bibitem{Yoshida:2015tia}
T.~Yoshida, E.~Hiyama, A.~Hosaka, M.~Oka and K.~Sadato,
\href{https://journals.aps.org/prd/abstract/10.1103/PhysRevD.92.114029}{Phys. Rev. D \textbf{92}, no.11, 114029 (2015)}
\href{https://arxiv.org/abs/1510.01067}{[arXiv:1510.01067 [hep-ph]]}.


\bibitem{Karliner:2015ema}
M.~Karliner and J.~L.~Rosner,
\href{https://journals.aps.org/prd/abstract/10.1103/PhysRevD.92.074026}{Phys. Rev. D \textbf{92}, no.7, 074026 (2015)}
\href{https://arxiv.org/abs/1506.01702}{[arXiv:1506.01702 [hep-ph]]}.


\bibitem{Thakkar:2016dna}
K.~Thakkar, Z.~Shah, A.~K.~Rai and P.~C.~Vinodkumar,
\href{https://arxiv.org/ct?url=https%3A%2F%2Fdx.doi.org%2F10.1016%2Fj.nuclphysa.2017.05.087&v=a0f0eee4}{Nucl. Phys. A \textbf{965}, 57-73 (2017)}
\href{https://arxiv.org/abs/1610.00411}{[arXiv:1610.00411 [nucl-th]]}.

\bibitem{Shah:2016mig}
Z.~Shah, K.~Thakkar, A.~Kumar Rai and P.~C.~Vinodkumar,
\href{https://link.springer.com/article/10.1140%2Fepja%2Fi2016-16313-9}{Eur. Phys. J. A \textbf{52}, no.10, 313 (2016)}
\href{https://arxiv.org/abs/1602.06384}{[arXiv:1602.06384 [hep-ph]]}.



\bibitem{Migura:2006ep}
S.~Migura, D.~Merten, B.~Metsch and H.~R.~Petry,
\href{https://link.springer.com/article/10.1140%2Fepja%2Fi2006-10017-9}{Eur. Phys. J. A \textbf{28}, 41 (2006)}
\href{https://arxiv.org/abs/hep-ph/0602153}{[arXiv:hep-ph/0602153 [hep-ph]]}.


\bibitem{Hussain:1999sp}
F.~Hussain, J.~G.~Korner and S.~Tawfiq,
\href{https://journals.aps.org/prd/abstract/10.1103/PhysRevD.61.114003}{Phys. Rev. D \textbf{61}, 114003 (2000)}
\href{https://arxiv.org/abs/hep-ph/9909278}{[arXiv:hep-ph/9909278 [hep-ph]]}.


\bibitem{Ivanov:1998wj}
M.~A.~Ivanov, J.~G.~Korner and V.~E.~Lyubovitskij,
\href{https://www.sciencedirect.com/science/article/pii/S0370269399000295?via%3Dihub}{Phys. Lett. B \textbf{448}, 143-151 (1999)}
\href{https://arxiv.org/abs/hep-ph/9811370}{[arXiv:hep-ph/9811370 [hep-ph]]}.


\bibitem{Ivanov:1999bk}
M.~A.~Ivanov, J.~G.~Korner, V.~E.~Lyubovitskij and A.~G.~Rusetsky,
\href{https://journals.aps.org/prd/abstract/10.1103/PhysRevD.60.094002}{Phys. Rev. D \textbf{60}, 094002 (1999)}
\href{https://arxiv.org/abs/hep-ph/9904421}{[arXiv:hep-ph/9904421 [hep-ph]]}.


\bibitem{Albertus:2005zy}
C.~Albertus, E.~Hernandez, J.~Nieves and J.~M.~Verde-Velasco,
\href{https://journals.aps.org/prd/abstract/10.1103/PhysRevD.72.094022}{Phys. Rev. D \textbf{72}, 094022 (2005)}
\href{https://arxiv.org/abs/hep-ph/0507256}{[arXiv:hep-ph/0507256 [hep-ph]]}.


\bibitem{Zhong:2007gp}
X.~H.~Zhong and Q.~Zhao,
\href{https://journals.aps.org/prd/abstract/10.1103/PhysRevD.77.074008}{Phys. Rev. D \textbf{77}, 074008 (2008)}
\href{https://arxiv.org/abs/0711.4645}{[arXiv:0711.4645 [hep-ph]]}.

\bibitem{Hernandez:2011tx}
E.~Hernandez and J.~Nieves,
\href{https://journals.aps.org/prd/abstract/10.1103/PhysRevD.84.057902}{Phys. Rev. D \textbf{84}, 057902 (2011)}
\href{https://arxiv.org/abs/1108.0259}{[arXiv:1108.0259 [hep-ph]]}.



\bibitem{Liu:2012sj}
L.~H.~Liu, L.~Y.~Xiao and X.~H.~Zhong,
\href{https://journals.aps.org/prd/abstract/10.1103/PhysRevD.86.034024}{Phys. Rev. D \textbf{86}, 034024 (2012)}
\href{https://arxiv.org/abs/1205.2943}{[arXiv:1205.2943 [hep-ph]]}.


\bibitem{Chen:2016iyi}
B.~Chen, K.~W.~Wei, X.~Liu and T.~Matsuki,
\href{https://link.springer.com/article/10.1140%2Fepjc%2Fs10052-017-4708-x}{Eur. Phys. J. C \textbf{77}, no.3, 154 (2017)}
\href{https://arxiv.org/abs/1609.07967}{[arXiv:1609.07967 [hep-ph]]}.


\bibitem{Wang:2017kfr}
K.~L.~Wang, Y.~X.~Yao, X.~H.~Zhong and Q.~Zhao,
\href{https://journals.aps.org/prd/abstract/10.1103/PhysRevD.96.116016}{Phys. Rev. D \textbf{96}, no.11, 116016 (2017)}
\href{https://arxiv.org/abs/1709.04268}{[arXiv:1709.04268 [hep-ph]]}.


\bibitem{Nagahiro:2016nsx}
H.~Nagahiro, S.~Yasui, A.~Hosaka, M.~Oka and H.~Noumi,
\href{https://journals.aps.org/prd/abstract/10.1103/PhysRevD.95.014023}{Phys. Rev. D \textbf{95}, no.1, 014023 (2017)}
\href{https://arxiv.org/abs/1609.01085}{[arXiv:1609.01085 [hep-ph]]}.


\bibitem{Yao:2018jmc}
Y.~X.~Yao, K.~L.~Wang and X.~H.~Zhong,
\href{https://journals.aps.org/prd/abstract/10.1103/PhysRevD.98.076015}{Phys. Rev. D \textbf{98}, no.7, 076015 (2018)}
\href{https://arxiv.org/abs/1803.00364}{[arXiv:1803.00364 [hep-ph]]}.






\bibitem{Wang:2018fjm}
K.~L.~Wang, Q.~F.~L\"u and X.~H.~Zhong,
\href{https://journals.aps.org/prd/abstract/10.1103/PhysRevD.99.014011}{Phys. Rev. D \textbf{99}, no.1, 014011 (2019)}
\href{https://arxiv.org/abs/1810.02205}{[arXiv:1810.02205 [hep-ph]]}.


\bibitem{Chen:2018vuc}
B.~Chen and X.~Liu,
\href{https://journals.aps.org/prd/abstract/10.1103/PhysRevD.98.074032}{Phys. Rev. D \textbf{98}, no.7, 074032 (2018)}
\href{https://arxiv.org/abs/1810.00389}{[arXiv:1810.00389 [hep-ph]]}.


\bibitem{Wang:2019uaj}
K.~L.~Wang, Q.~F.~L\"u and X.~H.~Zhong,
\href{https://journals.aps.org/prd/abstract/10.1103/PhysRevD.100.114035}{Phys. Rev. D \textbf{100}, no.11, 114035 (2019)}
\href{https://arxiv.org/abs/1908.04622}{[arXiv:1908.04622 [hep-ph]]}.




\bibitem{Zhu:2000py}
S.~L.~Zhu,
\href{https://journals.aps.org/prd/abstract/10.1103/PhysRevD.61.114019}{Phys. Rev. D \textbf{61}, 114019 (2000)}
\href{https://arxiv.org/abs/hep-ph/0002023}{[arXiv:hep-ph/0002023 [hep-ph]]}.

\bibitem{Wang:2010it}
Z.~G.~Wang,
\href{https://link.springer.com/article/10.1140%2Fepja%2Fi2011-11081-8}{Eur. Phys. J. A \textbf{47}, 81 (2011)}
\href{https://arxiv.org/abs/1003.2838}{[arXiv:1003.2838 [hep-ph]]}.


\bibitem{Mao:2015gya}
Q.~Mao, H.~X.~Chen, W.~Chen, A.~Hosaka, X.~Liu and S.~L.~Zhu,
\href{https://journals.aps.org/prd/abstract/10.1103/PhysRevD.92.114007}{Phys. Rev. D \textbf{92}, no.11, 114007 (2015)}
\href{https://arxiv.org/abs/1510.05267}{[arXiv:1510.05267 [hep-ph]]}.

\bibitem{Chen:2016phw}
H.~X.~Chen, Q.~Mao, A.~Hosaka, X.~Liu and S.~L.~Zhu,
\href{https://journals.aps.org/prd/abstract/10.1103/PhysRevD.94.114016}{Phys. Rev. D \textbf{94}, no.11, 114016 (2016)}
\href{https://arxiv.org/abs/1611.02677}{[arXiv:1611.02677 [hep-ph]]}.


\bibitem{Wang:2017vtv}
Z.~G.~Wang,
\href{https://www.sciencedirect.com/science/article/pii/S0550321317303814?via%3Dihub}{Nucl. Phys. B \textbf{926}, 467-490 (2018)}
\href{https://arxiv.org/abs/1705.07745}{[arXiv:1705.07745 [hep-ph]]}.


\bibitem{Mao:2017wbz}
Q.~Mao, H.~X.~Chen, A.~Hosaka, X.~Liu and S.~L.~Zhu,
\href{https://journals.aps.org/prd/abstract/10.1103/PhysRevD.96.074021}{Phys. Rev. D \textbf{96}, no.7, 074021 (2017)}
\href{https://arxiv.org/abs/1707.03712}{[arXiv:1707.03712 [hep-ph]]}.


\bibitem{Aliev:2018lcs}
T.~M.~Aliev, K.~Azizi, Y.~Sarac and H.~Sundu,
\href{https://journals.aps.org/prd/abstract/10.1103/PhysRevD.98.094014}{Phys. Rev. D \textbf{98}, no.9, 094014 (2018)}
\href{https://arxiv.org/abs/1808.08032}{[arXiv:1808.08032 [hep-ph]]}.


\bibitem{Cui:2019dzj}
E.~L.~Cui, H.~M.~Yang, H.~X.~Chen and A.~Hosaka,
\href{https://journals.aps.org/prd/abstract/10.1103/PhysRevD.99.094021}{Phys. Rev. D \textbf{99}, no.9, 094021 (2019)}
\href{https://arxiv.org/abs/1903.10369}{[arXiv:1903.10369 [hep-ph]]}.


\bibitem{Azizi:2020tgh}
K.~Azizi, Y.~Sarac and H.~Sundu,
\href{https://journals.aps.org/prd/abstract/10.1103/PhysRevD.101.074026}{Phys. Rev. D \textbf{101}, no.7, 074026 (2020)}
\href{https://arxiv.org/abs/2001.04953}{[arXiv:2001.04953 [hep-ph]]}.




\bibitem{Zhu:1998ih}
S.~L.~Zhu and Y.~B.~Dai,
\href{https://journals.aps.org/prd/abstract/10.1103/PhysRevD.59.114015}{Phys. Rev. D \textbf{59}, 114015 (1999)}
\href{https://arxiv.org/abs/hep-ph/9810243}{[arXiv:hep-ph/9810243 [hep-ph]]}.



\bibitem{Wang:2009cd}
Z.~G.~Wang,
\href{https://journals.aps.org/prd/abstract/10.1103/PhysRevD.81.036002}{Phys. Rev. D \textbf{81}, 036002 (2010)}
\href{https://arxiv.org/abs/0909.4144}{[arXiv:0909.4144 [hep-ph]]}.


\bibitem{Wang:2009ic}
Z.~G.~Wang,
\href{https://link.springer.com/article/10.1140%2Fepja%2Fi2010-10952-8}{Eur. Phys. J. A \textbf{44}, 105-117 (2010)}
\href{https://arxiv.org/abs/0910.2112}{[arXiv:0910.2112 [hep-ph]]}.


\bibitem{Aliev:2009jt}
T.~M.~Aliev, K.~Azizi and A.~Ozpineci,
\href{https://journals.aps.org/prd/abstract/10.1103/PhysRevD.79.056005}{Phys. Rev. D \textbf{79}, 056005 (2009)}
\href{https://arxiv.org/abs/0901.0076}{[arXiv:0901.0076 [hep-ph]]}.

\bibitem{Aliev:2010yx}
T.~M.~Aliev, K.~Azizi and M.~Savci,
\href{https://arxiv.org/ct?url=https%3A%2F%2Fdx.doi.org%2F10.1016%2Fj.physletb.2010.12.027&v=bb6d5acf}{Phys. Lett. B \textbf{696}, 220-226 (2011)}
\href{https://arxiv.org/abs/1009.3658}{[arXiv:1009.3658 [hep-ph]]}.

\bibitem{Aliev:2012tb}
T.~M.~Aliev, K.~Azizi and M.~Savci,
\href{https://aip.scitation.org/doi/abs/10.1063/1.4763508}{AIP Conf. Proc. \textbf{1492} (2012) no.1, 146-152}.

\bibitem{Aliev:2014bma}
T.~M.~Aliev, K.~Azizi and H.~Sundu,
\href{https://link.springer.com/article/10.1140%2Fepjc%2Fs10052-014-3229-0}{Eur. Phys. J. C \textbf{75}, no.1, 14 (2015)}
\href{https://arxiv.org/abs/1409.7577}{[arXiv:1409.7577 [hep-ph]]}.


\bibitem{Aliev:2016xvq}
T.~M.~Aliev, T.~Barakat and M.~Savc\i{},
\href{https://journals.aps.org/prd/abstract/10.1103/PhysRevD.93.056007}{Phys. Rev. D \textbf{93}, no.5, 056007 (2016)}
\href{https://arxiv.org/abs/1603.04762}{[arXiv:1603.04762 [hep-ph]]}.

\bibitem{Chen:2017sci}
H.~X.~Chen, Q.~Mao, W.~Chen, A.~Hosaka, X.~Liu and S.~L.~Zhu,
\href{https://journals.aps.org/prd/abstract/10.1103/PhysRevD.95.094008}{Phys. Rev. D \textbf{95}, no.9, 094008 (2017)}
\href{https://arxiv.org/abs/1703.07703}{[arXiv:1703.07703 [hep-ph]]}.


\bibitem{Agaev:2017ywp}
S.~S.~Agaev, K.~Azizi and H.~Sundu,
\href{https://journals.aps.org/prd/abstract/10.1103/PhysRevD.96.094011}{Phys. Rev. D \textbf{96}, no.9, 094011 (2017)}
\href{https://arxiv.org/abs/1708.07348}{[arXiv:1708.07348 [hep-ph]]}.

\bibitem{Aliev:2018vye}
T.~M.~Aliev, K.~Azizi, Y.~Sarac and H.~Sundu,
\href{https://journals.aps.org/prd/abstract/10.1103/PhysRevD.99.094003}{Phys. Rev. D \textbf{99}, no.9, 094003 (2019)}
\href{https://arxiv.org/abs/1811.05686}{[arXiv:1811.05686 [hep-ph]]}.


\bibitem{Chen:2007xf}
C.~Chen, X.~L.~Chen, X.~Liu, W.~Z.~Deng and S.~L.~Zhu,
\href{https://journals.aps.org/prd/abstract/10.1103/PhysRevD.75.094017}{Phys. Rev. D \textbf{75}, 094017 (2007)}
\href{https://arxiv.org/abs/0704.0075}{[arXiv:0704.0075 [hep-ph]]}.


\bibitem{Ye:2017yvl}
D.~D.~Ye, Z.~Zhao and A.~Zhang,
\href{https://journals.aps.org/prd/abstract/10.1103/PhysRevD.96.114009}{Phys. Rev. D \textbf{96}, no.11, 114009 (2017)}
\href{https://arxiv.org/abs/1709.00689}{[arXiv:1709.00689 [hep-ph]]}.

\bibitem{Ye:2017dra}
D.~D.~Ye, Z.~Zhao and A.~Zhang,
\href{https://journals.aps.org/prd/abstract/10.1103/PhysRevD.96.114003}{Phys. Rev. D \textbf{96}, no.11, 114003 (2017)}
\href{https://arxiv.org/abs/1710.10165}{[arXiv:1710.10165 [hep-ph]]}.


\bibitem{Chen:2017aqm}
B.~Chen, X.~Liu and A.~Zhang,
\href{https://journals.aps.org/prd/abstract/10.1103/PhysRevD.95.074022}{Phys. Rev. D \textbf{95}, no.7, 074022 (2017)}
\href{https://arxiv.org/abs/1702.04106}{[arXiv:1702.04106 [hep-ph]]}.

\bibitem{Yang:2018lzg}
P.~Yang, J.~J.~Guo and A.~Zhang,
\href{https://journals.aps.org/prd/abstract/10.1103/PhysRevD.99.034018}{Phys. Rev. D \textbf{99}, no.3, 034018 (2019)}
\href{https://arxiv.org/abs/1810.06947}{[arXiv:1810.06947 [hep-ph]]}.


\bibitem{Guo:2019ytq}
J.~J.~Guo, P.~Yang and A.~Zhang,
\href{https://journals.aps.org/prd/abstract/10.1103/PhysRevD.100.014001}{Phys. Rev. D \textbf{100}, no.1, 014001 (2019)}
\href{https://arxiv.org/abs/1902.07488}{[arXiv:1902.07488 [hep-ph]]}.


\bibitem{Liang:2019aag}
W.~Liang, Q.~F.~L\"u and X.~H.~Zhong,
\href{https://journals.aps.org/prd/abstract/10.1103/PhysRevD.100.054013}{Phys. Rev. D \textbf{100}, no.5, 054013 (2019)}
\href{https://arxiv.org/abs/1908.00223}{[arXiv:1908.00223 [hep-ph]]}.


\bibitem{Lu:2019rtg}
Q.~F.~L\"u and X.~H.~Zhong,
\href{https://journals.aps.org/prd/abstract/10.1103/PhysRevD.101.014017}{Phys. Rev. D \textbf{101}, no.1, 014017 (2020)}
\href{https://arxiv.org/abs/1910.06126}{[arXiv:1910.06126 [hep-ph]]}.


\bibitem{Padmanath:2013bla}
M.~Padmanath, R.~G.~Edwards, N.~Mathur and M.~Peardon,
\href{https://arxiv.org/abs/1311.4806}{[arXiv:1311.4806 [hep-lat]]}.



\bibitem{Bahtiyar:2015sga}
H.~Bahtiyar, K.~U.~Can, G.~Erkol and M.~Oka,
\href{https://arxiv.org/ct?url=https%3A%2F%2Fdx.doi.org%2F10.1016%2Fj.physletb.2015.06.006&v=b2ef763b}{Phys. Lett. B \textbf{747}, 281-286 (2015)}
\href{https://arxiv.org/abs/1503.07361}{[arXiv:1503.07361 [hep-lat]]}.


\bibitem{Bali:2015lka}
P.~P\'erez-Rubio, S.~Collins and G.~S.~Bali,
\href{https://journals.aps.org/prd/abstract/10.1103/PhysRevD.92.034504}{Phys. Rev. D \textbf{92}, no.3, 034504 (2015)}
\href{https://arxiv.org/abs/1503.08440}{[arXiv:1503.08440 [hep-lat]]}.


\bibitem{Bahtiyar:2016dom}
H.~Bahtiyar, K.~U.~Can, G.~Erkol, M.~Oka and T.~T.~Takahashi,
\href{https://www.sciencedirect.com/science/article/pii/S0370269317304896?via%3Dihub}{Phys. Lett. B \textbf{772}, 121-126 (2017)}
\href{https://arxiv.org/abs/1612.05722}{[arXiv:1612.05722 [hep-lat]]}.


\bibitem{Chow:1995nw}
C.~K.~Chow,
\href{https://journals.aps.org/prd/abstract/10.1103/PhysRevD.54.3374}{Phys. Rev. D \textbf{54}, 3374-3376 (1996)}
\href{https://arxiv.org/abs/hep-ph/9510421}{[arXiv:hep-ph/9510421 [hep-ph]]}.


\bibitem{Klempt:2009pi}
E.~Klempt and J.~M.~Richard,
\href{https://journals.aps.org/rmp/abstract/10.1103/RevModPhys.82.1095}{Rev. Mod. Phys. \textbf{82}, 1095-1153 (2010)}
\href{https://arxiv.org/abs/0901.2055}{[arXiv:0901.2055 [hep-ph]]}.


\bibitem{Crede:2013kia}
V.~Crede and W.~Roberts,
\href{https://iopscience.iop.org/article/10.1088/0034-4885/76/7/076301}{Rept. Prog. Phys. \textbf{76}, 076301 (2013)}
\href{https://arxiv.org/abs/1302.7299}{[arXiv:1302.7299 [nucl-ex]]}.


\bibitem{Chen:2016spr}
H.~X.~Chen, W.~Chen, X.~Liu, Y.~R.~Liu and S.~L.~Zhu,
\href{https://iopscience.iop.org/article/10.1088/1361-6633/aa6420}{Rept. Prog. Phys. \textbf{80}, no.7, 076201 (2017)}
\href{https://arxiv.org/abs/1609.08928}{[arXiv:1609.08928 [hep-ph]]}.


\bibitem{Aaij:2020fxj}
R.~Aaij \textit{et al.} [LHCb],
\href{https://arxiv.org/abs/2010.14485}{[arXiv:2010.14485 [hep-ex]]}.


\bibitem{Chen:2018orb}
B.~Chen, K.~W.~Wei, X.~Liu and A.~Zhang,
\href{https://journals.aps.org/prd/abstract/10.1103/PhysRevD.98.031502}{Phys. Rev. D \textbf{98}, no.3, 031502 (2018)}
\href{https://arxiv.org/abs/1805.10826}{[arXiv:1805.10826 [hep-ph]]}.


\bibitem{Yang:2019cvw}
H.~M.~Yang, H.~X.~Chen, E.~L.~Cui, A.~Hosaka and Q.~Mao,
\href{https://link.springer.com/article/10.1140%2Fepjc%2Fs10052-020-7637-z}{Eur. Phys. J. C \textbf{80}, no.2, 80 (2020)}
\href{https://arxiv.org/abs/1909.13575}{[arXiv:1909.13575 [hep-ph]]}.


\bibitem{Yang:2020zrh}
H.~M.~Yang and H.~X.~Chen,
\href{https://journals.aps.org/prd/abstract/10.1103/PhysRevD.101.114013}{Phys. Rev. D \textbf{101}, no.11, 114013 (2020)}
\href{https://journals.aps.org/prd/abstract/10.1103/PhysRevD.102.079901}{[erratum: Phys. Rev. D \textbf{102}, no.7, 079901 (2020)]}
\href{https://arxiv.org/abs/2003.07488}{[arXiv:2003.07488 [hep-ph]]}.


\bibitem{Faustov:2020gun}
R.~N.~Faustov and V.~O.~Galkin,
\href{https://www.mdpi.com/2571-712X/3/1/19}{Particles \textbf{3}, no.1, 234-244 (2020)}.

\bibitem{Faustov:2018vgl}
R.~N.~Faustov and V.~O.~Galkin,
\href{https://www.epj-conferences.org/articles/epjconf/abs/2019/09/epjconf_ishepp2019_08001/epjconf_ishepp2019_08001.html}{EPJ Web Conf. \textbf{204}, 08001 (2019)}
\href{https://arxiv.org/abs/1811.02232}{[arXiv:1811.02232 [hep-ph]]}.

\bibitem{Jia:2019bkr}
D.~Jia, W.~N.~Liu and A.~Hosaka,
\href{https://journals.aps.org/prd/abstract/10.1103/PhysRevD.101.034016}{Phys. Rev. D \textbf{101}, no.3, 034016 (2020)}
\href{https://arxiv.org/abs/1907.04958}{[arXiv:1907.04958 [hep-ph]]}.



\bibitem{Yu:2018yxl}
Q.~X.~Yu, R.~Pavao, V.~R.~Debastiani and E.~Oset,
\href{https://link.springer.com/article/10.1140%2Fepjc%2Fs10052-019-6665-z}{Eur. Phys. J. C \textbf{79}, no.2, 167 (2019)}
\href{https://arxiv.org/abs/1811.11738}{[arXiv:1811.11738 [hep-ph]]}.


\bibitem{Huang:2018bed}
Y.~Huang, C.~j.~Xiao, L.~S.~Geng and J.~He,
\href{https://journals.aps.org/prd/abstract/10.1103/PhysRevD.99.014008}{Phys. Rev. D \textbf{99}, no.1, 014008 (2019)}
\href{https://arxiv.org/abs/1811.10769}{[arXiv:1811.10769 [hep-ph]]}.



\bibitem{Nieves:2019jhp}
J.~Nieves, R.~Pavao and L.~Tolos,
\href{https://link.springer.com/article/10.1140%2Fepjc%2Fs10052-019-7568-8}{Eur. Phys. J. C \textbf{80}, no.1, 22 (2020)}
\href{https://arxiv.org/abs/1911.06089}{[arXiv:1911.06089 [hep-ph]]}.


\bibitem{Zhu:2020lza}
H.~Zhu and Y.~Huang,
\href{https://iopscience.iop.org/article/10.1088/1674-1137/44/8/083101}{Chin. Phys. C \textbf{44}, no.8, 083101 (2020)}
\href{https://arxiv.org/abs/2004.00728}{[arXiv:2004.00728 [hep-ph]]}.


\bibitem{Shifman:1978bx}
  M.~A.~Shifman, A.~I.~Vainshtein and V.~I.~Zakharov,
 \href{https://doi.org/10.1016/0550-3213(79)90022-1}{ Nucl.\ Phys.\ B {\bf 147}, 385 (1979)}.
  

  
\bibitem{Shifman:1978by}
  M.~A.~Shifman, A.~I.~Vainshtein and V.~I.~Zakharov,
 \href{https://doi.org/10.1016/0550-3213(79)90023-3}{ Nucl.\ Phys.\ B {\bf 147}, 448 (1979)}.
 
 

 \bibitem{Ioffe81} 
  B.~L.~Ioffe,
 \href{https://doi.org/10.1016/0550-3213(81)90259-5}{ Nucl.\ Phys.\ B {\bf 188}, 317 (1981)}
 \href{https://doi.org/10.1016/0550-3213(81)90315-1}{ Erratum: [Nucl.\ Phys.\ B {\bf 191}, 591 (1981)]}.






\bibitem{Braun:1988qv} 
  V.~M.~Braun and I.~E.~Filyanov,
\href{https://link.springer.com/article/10.1007/BF01548594}{  Z.\ Phys.\ C {\bf 44}, 157 (1989)}
  [Sov.\ J.\ Nucl.\ Phys.\  {\bf 50}, 511 (1989)]
  [Yad.\ Fiz.\  {\bf 50}, 818 (1989)].
  
  
  

\bibitem{Balitsky:1989ry} 
  I.~I.~Balitsky, V.~M.~Braun and A.~V.~Kolesnichenko,
\href{https://www.sciencedirect.com/science/article/abs/pii/0550321389905701?via%3Dihub}{  Nucl.\ Phys.\ B {\bf 312}, 509 (1989)}.

\bibitem{Chernyak:1990ag} 
  V.~L.~Chernyak and I.~R.~Zhitnitsky,
\href{https://www.sciencedirect.com/science/article/abs/pii/055032139090612H?via%3Dihub}{  Nucl.\ Phys.\ B {\bf 345}, 137 (1990)}.
  
  


\bibitem{Zyla:2020zbs}
P.~A.~Zyla \textit{et al.} [Particle Data Group],
\href{https://academic.oup.com/ptep/article/2020/8/083C01/5891211}{PTEP \textbf{2020}, no.8, 083C01 (2020)}.
  
  
  

 \bibitem{Azizi:2016dmr} 
  K.~Azizi, N.~Er and H.~Sundu,
\href{https://www.sciencedirect.com/science/article/pii/S0375947417300325?via%3Dihub}{  Nucl.\ Phys.\ A {\bf 960}, 147 (2017)}
  Erratum: [Nucl.\ Phys.\ A {\bf 962}, 122 (2017)]
\href{https://arxiv.org/abs/1605.05535}{  [arXiv:1605.05535 [hep-ph]]}.

\bibitem{Belyaev:1982sa}
  V.~M.~Belyaev and B.~L.~Ioffe,
  Sov. Phys. JETP \textbf{56}, 493-501 (1982)
 \href{http://www.jetp.ac.ru/cgi-bin/dn/e_056_03_0493.pdf}{ [Zh.\ Eksp.\ Teor.\ Fiz.\  {\bf 83}, 876 (1982)]}.
  
  

  
   \bibitem{Belyaev:1982cd}
  V.~M.~Belyaev and B.~L.~Ioffe,
 Sov.\ Phys.\ JETP {\bf 57}, 716 (1983)
   \href{http://www.jetp.ac.ru/cgi-bin/dn/e_057_04_0716.pdf}{[Zh.\ Eksp.\ Teor.\ Fiz.\  {\bf 84}, 1236 (1983)]}.






  
  
 

\bibitem{Chetyrkin:2007vm}
  K.~G.~Chetyrkin, A.~Khodjamirian and A.~A.~Pivovarov,
 \href{https://www.sciencedirect.com/science/article/pii/S0370269308002360?via%3Dihub}{ Phys.\ Lett.\ B {\bf 661} (2008) 250}
 \href{https://arxiv.org/abs/0712.2999}{ [arXiv:0712.2999 [hep-ph]]}.

\bibitem{Aliev:2018ube}
T.~M.~Aliev, K.~Azizi and H.~Sundu,
 \href{https://doi.org/10.1140/epja/i2018-12593-3}{Eur. Phys. J. A \textbf{54}, no.9, 159 (2018)}
 \href{https://arxiv.org/abs/1803.04002}{[arXiv:1803.04002 [hep-ph]]}.




\bibitem{Ball:2006wn}
  P.~Ball, V.~M.~Braun and A.~Lenz,
  \href{https://iopscience.iop.org/article/10.1088/1126-6708/2006/05/004}{JHEP {\bf 0605}, 004 (2006)}
  \href{https://arxiv.org/abs/hep-ph/0603063}{[hep-ph/0603063]}.
  
  
  
  

\bibitem{Belyaev:1994zk}
  V.~M.~Belyaev, V.~M.~Braun, A.~Khodjamirian and R.~Ruckl,
\href{https://journals.aps.org/prd/abstract/10.1103/PhysRevD.51.6177}{  Phys.\ Rev.\ D {\bf 51}, 6177 (1995)}
\href{https://arxiv.org/abs/hep-ph/9410280}{  [hep-ph/9410280]}.



\bibitem{Ball:2004ye}
  P.~Ball and R.~Zwicky,
\href{https://journals.aps.org/prd/abstract/10.1103/PhysRevD.71.014015}{  Phys.\ Rev.\ D {\bf 71}, 014015 (2005)}
 \href{https://arxiv.org/abs/hep-ph/0406232}{[hep-ph/0406232]}.



\bibitem{Ball:2004hn}
P.~Ball and R.~Zwicky,
\href{https://arxiv.org/abs/hep-ph/0406261}{[arXiv:hep-ph/0406261 [hep-ph]]}.











\end{thebibliography}
\end{document}